\documentclass[aps, prx, amsmath, amssymb, amsfonts, superscriptaddress, floatfix, twocolumn, 10pt, longbibliography]{revtex4-2}

\usepackage[usenames, dvipsnames]{color}
\usepackage[dvipsnames,svg justnames,x11names,hyperref]{xcolor}

\usepackage[T1]{fontenc}
\usepackage[utf8]{inputenc}
\usepackage{times}
\usepackage[low-sup]{subdepth}
\usepackage{float}
\usepackage{graphicx}
\usepackage{tabularx}
\newcolumntype{R}{>{\raggedleft\arraybackslash}X}
\usepackage{ragged2e}
\usepackage{enumerate}

\usepackage[
]{physpack}

\usepackage{pdfpages}
\makeatletter
\AtBeginDocument{\let\LS@rot\@undefined}
\makeatother

\makeatletter
\newcommand*{\balancecolsandclearpage}{%
  \close@column@grid
  \clearpage
  \twocolumngrid
}
\makeatother

\definecolor{linkcolor}{RGB}{6,69,173} % Wikipedia
\usepackage[
    unicode=true,
    pdfusetitle,
    bookmarks=false,
    colorlinks=true,
    linkcolor=linkcolor,
    urlcolor=linkcolor,
    citecolor=linkcolor,
    pdfencoding=auto]{hyperref}

\makeatletter
\@ifundefined{date}{}{\date{}}
\AtBeginDocument{
  
}
\makeatother

\renewcommand{\bra}[1]{\left\langle{#1}\right|}
\renewcommand{\ket}[1]{\left|{#1}\right\rangle}

\DeclareMathOperator{\sgn}{sgn}
\DeclareMathOperator{\Tr}{Tr}
\DeclareMathOperator{\SWAP}{SWAP}

\begin{document}

\title{
%Floquet generation of strong zero mode
Emergent strong zero mode through local Floquet engineering}

\author{Bhaskar Mukherjee}
\affiliation{Department of Physics and Astronomy, University College London, Gower Street, London WC1E 6BT, United Kingdom}

\author{Ronald Melendrez}
\affiliation{Department of Physics, Florida State University, Tallahassee, Florida 32306, USA}
\affiliation{National High Magnetic Field Laboratory, Tallahassee, Florida 32304, USA}

\author{Marcin Szyniszewski}
\affiliation{Department of Physics and Astronomy, University College London, Gower Street, London WC1E 6BT, United Kingdom}

\author{Hitesh J. Changlani}
\affiliation{Department of Physics, Florida State University, Tallahassee, Florida 32306, USA}
\affiliation{National High Magnetic Field Laboratory, Tallahassee, Florida 32304, USA}

\author{Arijeet Pal}
\affiliation{Department of Physics and Astronomy, University College London, Gower Street, London WC1E 6BT, United Kingdom}

\date{\today}

\begin{abstract}
 Periodically driven quantum systems host exotic phenomena which often do not have any analog in  undriven systems. Floquet prethermalization and dynamical freezing of certain observables, via the emergence of conservation laws, are realized by controlling the drive frequency. These dynamical regimes can be leveraged to construct quantum memories and have potential applications in quantum information processing. Solid state and cold atom experimental architectures have opened avenues for implementing local Floquet engineering which can achieve spatially modulated quantum control of states. Here, we uncover the novel memory effects of local periodic driving in a nonintegrable spin-half staggered Heisenberg chain. For a boundary-driven protocol at the dynamical freezing frequency, we show the formation of an approximate strong zero mode, a prethermal quasi-local operator, due to the emergence of a discrete global $\mathbb{Z}_2$ symmetry. This is captured by constructing an accurate effective Floquet Hamiltonian using a higher-order partially resummed Floquet-Magnus expansion. The lifetime of the boundary spin can be exponentially enhanced by enlarging the set of suitably chosen driven sites. We demonstrate that in the asymptotic limit, achieved by increasing the number of driven sites, a strong zero mode emerges, where the lifetime of the boundary spin grows exponentially with system size. The non-local processes in the Floquet Hamiltonian play a pivotal role in the total freezing of the boundary spin in the thermodynamic limit. The novel dynamics of the boundary spin is accompanied by a rich structure of entanglement in the Floquet eigenstates where specific bipartitions yield an area-law scaling while the entanglement for random bipartitions scales as a volume-law.

\end{abstract}
\maketitle

\section{Introduction}
%\begin{itemize}
%\item Floquet systems (prethermalization, dynamic freezing, Floquet engineering)\\
%\item Global vs local driving\\
%\item scars\\
%\item SZM\\
%\end{itemize}

Protection of quantum information in many-body systems in the presence of decoherence is vital for quantum information processing~\cite{TerhalQEC,reviewtqc, sarma2015majorana}. Realizations in physical systems of qubits are routinely subjected to periodic drives using optical and microwave resonators for quantum control~\cite{majer2007coupling, girvin2014circuit, DuanMonroe_2010}. Driving systems away from equilibrium is an experimental prescription for lengthening the lifetime of qubits. In recent years Floquet systems have been shown to host novel states of matter characterized by dynamical localization~\cite{dl,dlexpt}, topology~\cite{rudner, rudner2020FTop, wintersperger2020_FTopo}, and time crystallinity~\cite{sacha2017TC, Else_DTC, khemani2019DTC, Else2017_Prethermal, Bluvstein, Masakara, Sullivan2020DTC, randall2021DTC}. Realizing exotic dynamical states through Floquet engineering has witnessed a surge in theoretical interest~\cite{moessner2017Floquet, oka2019floquet}. In interacting, driven systems thermalization to infinite temperature provides a major roadblock to the protection of entanglement and correlations~\cite{rigolprx,ad1}. In the absence of disorder, quantum many-body scars (QMBS)~\cite{turner2018Scars, serbyn2021QMBS, sanjay1, sanjay2, regnault2022QMBS, Lee_PRBR2020, klebanov,Jeyaretnam2021ScarredSPT, katsura} and Hilbert-space fragmentation~\cite{Sala2020_HSF, Khemani2020_HSF, Lee_Pal_Changlani_2021, sanjayPRX, minimalHSF, Richter2022ScarredFF} have provided instances where local thermalization can be avoided due to constraints in the interaction.

Certain one-dimensional many-body Hamiltonians are endowed with a quasi-local operator called a strong zero mode (SZM), which exists on the boundary of the system and is conserved for long times \cite{fendley,kemp}. This operator can be utilized to encode stable quantum information away from the ground state and does not require low temperatures. The existence of SZM can be proven in integrable models while in the presence of weak integrability-breaking perturbations, these quasi-local operators are expected to exhibit prethermal behavior \cite{else,aditi,aditi1}.  Models with Hilbert space fragmentation can host a statistically localized SZM, without the presence of integrability \cite{SLIOM}. The connection between integrability and SZMs is of significant interest for the breakdown of thermalization. In this context, symmetries play a crucial role in fragmenting the Hilbert space and forming quantum scars, which stabilize SZM. This class of phenomena falls under the rubric of partially integrable models, where local dynamical properties exhibit non-thermal behavior. Under a global periodic drive that breaks integrability while preserving the global symmetry, the SZM destabilizes and has a finite lifetime~\cite{Yates2019,Yates2021_FSZM}. These boundary observables take a parametrically long time to thermalize while the bulk of the system thermalizes quickly. These approximate zero modes are continuously connected to exact zero modes in the integrable limit of the models.  Indeed, in most cases, the presence of a global drive appears to be crucial for realizing prethermal or athermal behavior due to the emergence of conservation laws.

Generically, an interacting Floquet system heats up to an infinite temperature state, however, prethermal or slow dynamics can be induced due to the separation of scales between the frequency of the drive and the system's internal energy scales~\cite{Abanin1,Abanin2,kuwahara,DElse, AbadalPRX_Fprethermal}.  The distinct stroboscopic dynamical regimes of a periodically driven system are governed by the Floquet Hamiltonian $H_F$, defined through the propagator $U(\tau)$ as 
\begin{equation}
U(\tau) = \mathcal{T} \exp(-i\int_0^\tau H(t')dt')=\exp(-iH_F \tau),   
\end{equation}
where $H(t) = H(t+\tau)$. $H_F$ plays a central role in understanding the slow dynamics or heating to infinite temperature~\cite{anatoli1, BWoka, Eckardt_2015}.  
A perturbative description of the prethermal timescales can be developed in terms of an expansion of $H_F$ in $\tau$, where $\tau$ is the time-period of the Floquet drive. In general, the perturbative corrections in a periodically driven nonintegrable system become non-local and the series diverges in the thermodynamic limit, signaling rapid thermalization. Therefore, in finite size systems the choice of drive protocol which increases the radius of convergence of such perturbation series is of importance for Floquet engineering. In the presence of a high-frequency drive, nonintegrable systems can exhibit prethermal behavior where emergent conservation laws are applicable for exponentially long times. 
In another class of phenomena global drives can result in the formation of Floquet quantum scars~\cite{Mukherjee2020_Floquetscars, Shriya, bm, Ana, Michailidis2020_drivenscars,kartiek} and dynamical many-body freezing~\cite{ad,asmi, satyaki}, where local observables remain athermal under unitary evolution for a specific choice of drive protocols -- an exclusively drive-induced non-thermal phenomenon at intermediate frequencies. The frozen states are characterized by local conservation laws and lead to scarring of the Floquet spectrum. 

A question naturally emerges, what happens when a system is driven locally? Unlike the global drive which can preserve translational symmetry, a local drive explicitly breaks it. Recently, the real-space profile of thermalization for local or spatially modulated drives has been shown to exhibit rich behavior where the drive can disentangle spins leading to large variations in thermalization times and even steady-state properties~\cite{melendrez2022localdrive, wen2022floquetcooling}. The disentangling effect of the drive can generate \textit{cold} spots which persist for significantly longer than thermalization times in the rest of the system. 

\begin{figure}
    \centering
    \includegraphics[width=\linewidth]{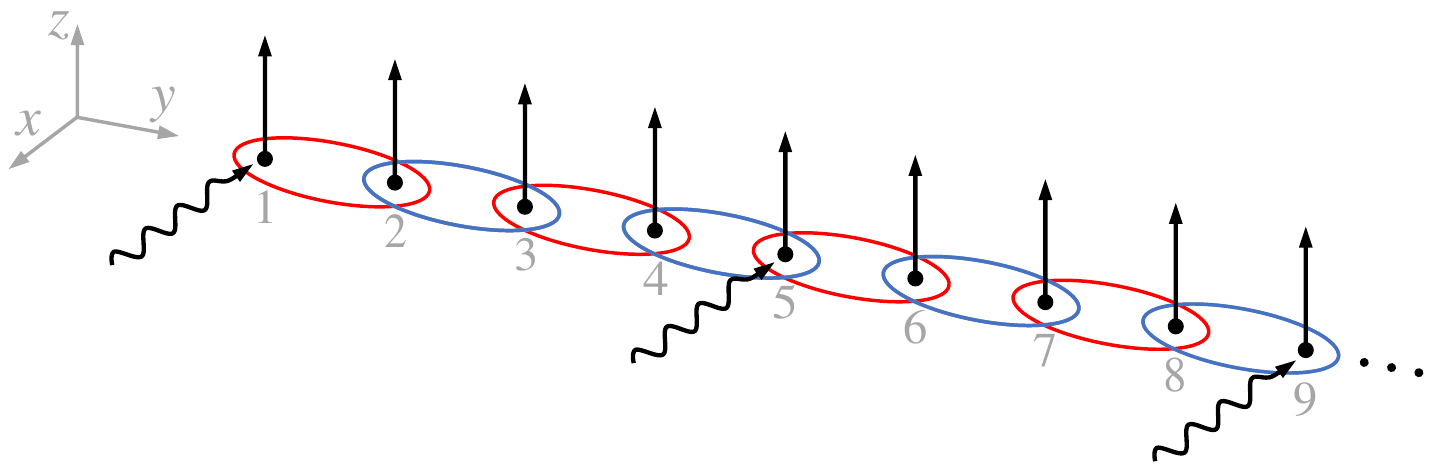}
    \caption{A schematic figure showing the model interaction and drive protocol adapted in this work. Red (blue) color is used to denote ferro- (antiferro-) magnetic exchange in consecutive links. A stationary global field is applied in the z-direction whereas a time-periodic field in the x-direction is applied at a few specific sites.}
    \label{fig:schematic}
\end{figure}

In this work, we consider the fate of thermalization for a boundary-driven nonintegrable spin chain. In the absence of driving, the model hosts a set of scarred eigenstates that act as a large spin given by the SU(2) symmetry of the model, showing persistent oscillations for certain unentangled initial states. In the presence of a local drive, this manifold is destroyed by a novel form of dynamical disentangling of the spin due to the emergence of a discrete symmetry at the dynamical freezing frequency. We unravel a new mechanism for local ergodicity breaking due to the formation of an approximate SZM at a certain discrete set of frequencies. Furthermore, using Floquet engineering of a cluster of sites (see the illustration in Fig.~\ref{fig:schematic}) we are able to control the exponential time scales of boundary relaxation characterized by the emergence of an asymptotically exact SZM. The athermal dynamics is accompanied by a rich entanglement structure of Floquet eigenstates which exhibit both thermal and athermal properties.

The rest of the paper is structured as follows. In Sec.~\ref{sec:model} we introduce our model, and discuss its symmetries and nonintegrable nature. This is followed by the definition of the local Floquet protocol. In Sec.~\ref{sec:singlesite} we discuss the boundary-driven protocol where the drive acts on a single site. Using higher-order Floquet-Magnus (F-M) expansion, we demonstrate the slow dynamics of the edge spin and the resulting properties of the approximate SZM. In Sec.~\ref{sec:multisite}, we give the full phenomenology of two and multi-site driving, discovering the optimal drive protocol which leads to the freezing of the boundary spin. We discuss how this novel dynamical property is accompanied by the emergence of a SZM in Sec.~\ref{sec:szm}. In order to interpret our results on the dynamics, we analyze the entanglement structures of the Floquet eigenstates in Sec.~\ref{sec:EE}, and show their thermal and athermal properties for specific bipartitions. Sec.~\ref{sec:discussion} summarises the salient results and their implications for athermal behavior in Floquet systems.

\section{Model Hamiltonian, drive protocol and preliminaries}
\label{sec:model}
%\begin{itemize}
%\item introduce the model, drive protocol, cite our previous work\\
%\end{itemize}

\subsection{Undriven Hamiltonian}
We consider the one-dimensional ``staggered'' (alternating ferro-antiferromagnetic exchange interaction) nearest neighbor spin-1/2 Heisenberg model with a globally applied homogeneous magnetic field, as the undriven part of our system. Its Hamiltonian is,
\begin{equation}
H_0=\sum_{i=1}^{N-1}(-1)^i{\bf S}_i \cdot {\bf S}_{i+1}-h\sum_{i=1}^NS^z_i,
\label{eq:model}
\end{equation}
where $i$ refers to a site index and ${\bf S}_i\equiv(S^x_i,S^y_i,S^z_i)$ is the vector comprised of the usual spin-1/2 operators, and $h$ is the magnetic field strength. %=\sigma^{\alpha}_i / 2$. 
$N$ is the number of spins and we always consider open boundary conditions (OBC). 
As previously mentioned elsewhere by some of us~\cite{melendrez2022localdrive}, the model in Eq.~\eqref{eq:model} has a long history partly due to its relevance to the Haldane spin-1 chain \cite{hida,hal_tasaki,hida2,compound}, which can be realized in the limit of the ferromagnetic bonds being much stronger than the antiferromagnetic ones.

For $h \neq 0 $, the magnetic field term reduces the symmetry of the Hamiltonian from SU(2) to $U(1)$, thus the total magnetization along the $z$ axis ($S^z_\text{tot}=\sum_iS^z_i$) is conserved. A subset of $U(1)$ is the $\mathbb{Z}_2$ symmetry,  i.e.\@ $H_0$ commutes with the $D_z$ operator, where
\begin{equation}
    D_\alpha = \prod_{i=1}^N 2S^\alpha_i
\end{equation}
with $\alpha = x, y, z$. In terms of spatial symmetries, $H_0$ is reflection symmetric (w.r.t.\@ the central bond) only for even $N$. For odd $N$, $H_0$ is found to anticommute with the operators 
\begin{equation}
        \mathcal{A}_\alpha \equiv D_\alpha R,
\end{equation}
for $\alpha=x,y$, where we have defined,
\begin{eqnarray}
    R&\equiv&\prod_{i=1}^{(N-1)/2}\SWAP(i,N-i+1).
\end{eqnarray}
$R$ is the reflection operator w.r.t.\@ the central site and $\SWAP(i,j)$ is the SWAP gate, defined as $\SWAP \equiv | {\uparrow} {\uparrow} \rangle \langle {\uparrow} {\uparrow} | + | {\uparrow} {\downarrow} \rangle \langle {\downarrow} {\uparrow} | + | {\downarrow} {\uparrow} \rangle \langle {\uparrow} {\downarrow} | + | {\downarrow} {\downarrow} \rangle \langle {\downarrow} {\downarrow} |$. Consequently, $H_0$ has a spectral reflection symmetry (i.e.\@ if there is an eigenstate $\ket{\psi}$ of energy $E$, then there is also an eigenstate $\mathcal{A}\ket{\psi}$ with energy $-E$) for odd $N$. 

We find that there are exponentially many (with system size) exact zero energy states at $h=0$ for odd $N$, intriguingly such a feature has also been reported in other models that harbor a superspin structure, which translates to the existence of QMBS \cite{turner2018Scars,turnerPRB,iadecolaZeromodes,senZeromodes,bmspin1pxp, Sharma_Lee_Changlani}. For any eigenstate $\ket{\psi}$ with magnetization $m_z$, there is always an eigenstate $R\ket{\psi}$ ($D_{\alpha} \ket{\psi}$, for $\alpha=x,y$) with magnetization $m_z$ ($-m_z$). For even $N$, both the spectral reflection ($E \rightarrow -E$) symmetry and exact zero energy states are lost but the latter property still remain valid.

\begin{figure}
    \centering
    \includegraphics[width=\linewidth]{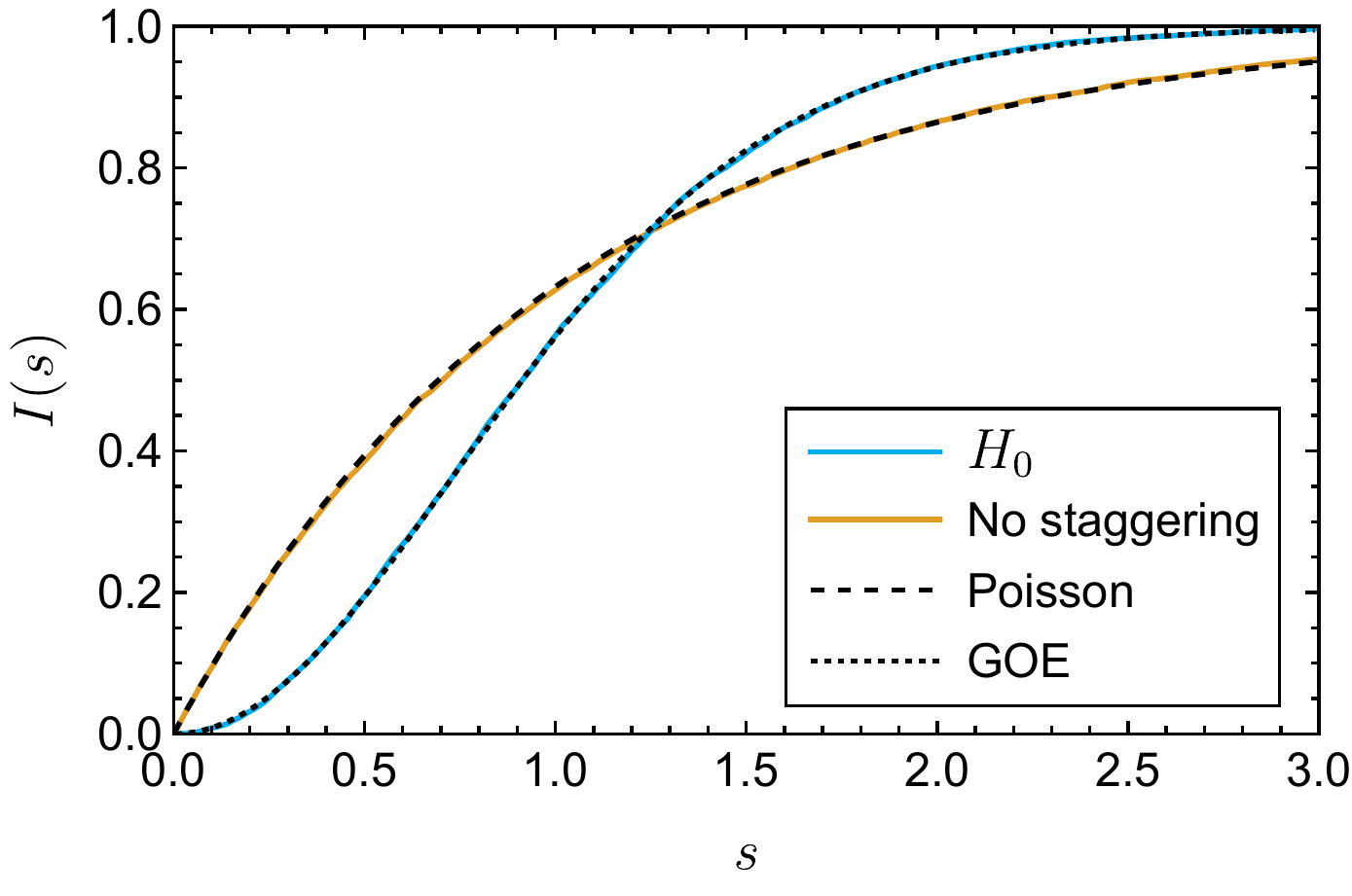}
    \caption{Integrated level spacing distribution $I(s)$ for $H_0$ (in blue) and $H_0$ without staggering (in yellow), both with $h = 1, N = 16$. Black lines show a comparison with Poisson statistics, the signature of integrability (dashed), and GOE, valid for chaotic systems (dotted).}
    \label{fig:levelstat}
\end{figure}

Despite its superficial resemblance to the integrable uniform Heisenberg chain, the staggered model is nonintegrable.
Focusing on even $N$, here $N=16$, we work in the zero-magnetization sector ($S^z_\text{tot}=0$) and switch off the magnetic field at the last site to remove the presence of conventional symmetries (e.g.\@ $S^2$ conservation, inversion symmetry)~\cite{Benenti2009}, which must otherwise be accounted for in the determination of level statistics. The middle 60\% of the many-body spectrum is used for the purpose of computing level statistics, this is done to focus on properties at infinite temperature.
Due to the limited statistics of eigenvalue spacings available for every bin, it is instead useful to monitor the integrated level spacing distribution,
\begin{equation}
I(s) = \int_0^{s} P(s') ds',
\end{equation}
where $P(s')$ is the probability density of the unfolded energy level spacings (spacings divided by their mean). Note that $I(s)$ is simply the cumulative distribution function of the unfolded spacings $s$.

In Fig.~\ref{fig:levelstat} we plot $I(s)$ for the case of $h=1$, and find that it closely resembles the results of a Gaussian orthogonal ensemble (GOE) of the random matrix theory~\cite{Haake}, a signature of chaotic, nonintegrable systems~\footnote{We find similar results for $h \rightarrow 0$ i.e.\@ even the staggered Heisenberg model without a field is nonintegrable, consistent with previous integrability tests of Ref.~\cite{Grabowski1995}.}. In comparison, when the bonds are made uniform (instead of staggered), we find that $I(s)$ follows the Poisson distribution, as expected for integrable models. 
%We have worked in the zero-magnetization sector ($S^z_\text{tot}=0$) and switched off the magnetic field at the last site to remove any further possible conventional symmetry (e.g.\@ $S^2$ conservation, inversion symmetry)~\cite{Benenti2009}; also, only the middle 60\% of the spectrum is used. 
%Note that all our conclusions, presented in the rest of the paper, do not %depend much on the presence (or absence) of the field at the last site.

\subsection{Strong zero mode}

Our work here is inspired by the aim of engineering a strong zero mode, a quasi-local conserved operator, localized at the edge of a one-dimensional quantum system which gives rise to long coherence times of the edge spin~\cite{fendley, kemp, else, kempPRL, SLIOM}. Given previous work on the search and characterization of such modes, it has been established that the SZM operator $\hat{O}_{\text{szm}}$ in a spin-1/2 system must satisfy the following conditions:
\begin{enumerate}
   \item There exists a discrete symmetry operator $D$ in the system, i.e.\@ $[D, H]=0$, implying the eigenstates are split into two sectors, labeled by the eigenvalues of $D$.
   
   \item $\hat{O}_{\text{szm}}$ anticommutes with $D$, $\{ \hat{O}_{\text{szm}}, D \} = 0$, and therefore pairs the eigenstates between the sectors of $D$.
   
   \item $\hat{O}_{\text{szm}}$ must be normalizable, $\hat{O}^2_{\text{szm}}\propto\mathbb{I}$.
   
   \item $\hat{O}_{\text{szm}}$ asymptotically commutes with the Hamiltonian $H$ of the system, i.e.\@ $||[H,\hat{O}_{\text{szm}}]||\sim\exp(-aN)$ with $a>0$. A consequence of this is that $\hat{O}_\text{szm}$ is a conserved quantity in the thermodynamic limit and that the gap falls off exponentially in the system size.
\end{enumerate}
We, therefore, expect that a system endowed with a SZM is characterized by a spectrum-wide (quasi)-degeneracy and relaxation of the edge that can be slowed down by increasing the system size. It is also important to satisfy all these conditions simultaneously to have SZM in a system, for example, one can construct an operator which satisfies the last condition but does not correspond to a SZM. The undriven staggered Heisenberg chain ($H_0$) is not expected to possess a normalizable SZM and, consequently, any edge spin should quickly relax to zero starting from any randomly chosen ordered state. In a bid to effectively decouple the edge spin from the rest of the system, we adopt the strategy of driving it locally with a protocol we describe next.

\subsection{Drive protocol}
Throughout this work, we adopt a Floquet drive protocol that involves applying a time-periodic magnetic field in the $x$-direction on a subsystem that includes the boundary spin. The total time-dependent Hamiltonian is given by
\begin{align}
    H(t)&=H_0+f(t)V \ ,\nonumber\\
  f(t)&=\gamma \sgn[\sin(\omega t)] \ \ \text{and} \ \ V=\sum_{i\in s_d}S^x_i,
    \label{eq:protocol}
\end{align}
where $\sgn(x)$ is the sign function, and thus its profile is that of a ``square pulse", and $s_d$ is the set of driven sites (see Fig.~\ref{fig:schematic}). $\gamma$ and $\omega$ are the drive amplitude and frequency, respectively and $\tau \equiv {2\pi}/{\omega}$ is the time duration of one drive cycle. The time-averaged field in the $x$-direction is zero at stroboscopic times (integer number of drive cycles).
We note that some aspects were previously considered for the special case of a single driven bulk site~\cite{Thuberg,adhip,Hubner}.

The time evolution operator over one drive cycle can be written as
\begin{equation}
    U(\tau)=e^{-i(H_0-\gamma V)\frac{\tau}{2}}e^{-i(H_0+\gamma V)\frac{\tau}{2}}=e^{-iH_F \tau}.
\end{equation}
%where the last equality follows from the Floquet theorem, and $H_F$ is defined to be the Floquet Hamiltonian. 
Since we focus on stroboscopic dynamics, $t=n\tau$ where $n$ is an integer, analyzing the properties of $U(\tau)$ or equivalently $H_F$ is adequate for long time dynamics. A key contribution of our work is to explicitly derive $H_F$, to some approximation, in a bid to understand the exact stroboscopic dynamics seen in our numerical calculations. The explicit form of $H_F$ was computed to the lowest order in the F-M expansion for the case of a single driven site in the bulk \cite{melendrez2022localdrive}, a result we revisit in the next section for a boundary driven case.

\subsection{Observables}
%We are interested in stroboscopic local dynamics of the system for which we will be studying the expectation value of local observables, 
Both from an experimental and theoretical point of view it is advantageous to monitor the time dependence of local operators, here we focus on $\langle \psi(n\tau) | S^x_i | \psi(n\tau)\rangle$, where
\begin{equation}
\ket{\psi(n\tau)} = U(n\tau) | \psi_0 \rangle = U^{n}(\tau) | \psi_0 \rangle
\end{equation}
is the time-dependent state starting from an initial state $|\psi_0 \rangle$. The initial state is typically chosen to be ``simple", for example, a product state in the $x$ basis, with the hope that it can be easily realized in experiments~\cite{Bernien2017,Bluvstein,Tan_TFIM_2021,Jepsen2022}. 

Further theoretical insights can be obtained from the time-dependent 
von Neumann entanglement entropy,
\begin{equation}
S^{\text{vN}}_A (n\tau)=\Tr[\rho_A(n\tau)\ln\rho_A (n\tau)],
\end{equation}
where $A$ refers to a single site or collection of sites. The reduced density matrix of this collection of sites, $\rho_A$, is computed by tracing out all degrees of freedom not part of $A$ (labeled by $\bar{A}$) in the full density matrix
\begin{equation}
\rho_A (n\tau)=\Tr_{\bar{A}}\Big[\ket{\psi(n\tau)}\bra{\psi(n\tau)}\Big].
\end{equation}
In this work, we choose $A$ to include either a single site, or a collection of driven sites. Entanglement entropy is a useful metric for distinguishing between low-entangled states (area law or lower) or those with high entanglement (typically, volume law). Thus, it reveals the existence or absence of (Floquet) ETH and is also useful for the detection of QMBS~\cite{turner2018Scars, Lee_PRBR2020}.
%the fully x-polarized initial state: $\ket{\psi(nT)}=U(nT)\ket{\psi_0}$ where $\ket{\psi_0}=\bigotimes_i\ket{\leftarrow}_i$ and $S^x_i\ket{\leftarrow}_i=-\frac{1}{2}\ket{\leftarrow}_i$ (we assume $\hbar=1$). $\ket{\psi_0}$ can be decomposed in the $S^z_\text{tot}$ basis as 

\section{Boundary-driven approximately-conserved strong zero mode}
\label{sec:singlesite}

\subsection{Overview of salient features of dynamics}
Given the setup in the previous section, we study the stroboscopic relaxation of the spins when only the left boundary spin  is driven i.e.\@ $s_d=\{1\}$.
The main purpose is to explore the existence of parameters that freeze the dynamics of the edge spin via the emergence of a conserved quantity (for example, $S^x_1$) which can lead to realizing a strong SZM.
Instead, we find  an \textit{approximate} SZM for the boundary-driven protocol, i.e., no operator satisfies the criteria outlined previously. However, the lessons learned from understanding this case facilitate a series of drive protocols that lead us to a SZM, demonstrated in the following sections.

We summarize our key physical findings in this section, which relies on constructing $H_F$ using a resummed F-M expansion to different levels of accuracy. %(as charted out in detail in Appendix~\ref{app:A}). 
%Thus, $H_F=\sum_{l=0}^{\infty}H_F^{(l)}$ was calculated term by term.
First, we revisit the result for $H_F$ to zeroth order ($l=0$) in the F-M expansion, by transforming to the rotating frame. We show that this approximation, while qualitatively correct for many situations, fails to capture the dynamics evaluated from exact numerics near certain special drive frequencies which we dub as ``freezing condition''\cite{asmi}. This failure motivates the technically challenging computation of $H_F$ to second order which contains multiple terms, we classify as ``local" and ``non-local". We find that the local terms, treated accurately to high order in the inverse drive strength, renormalize the freezing frequency in excellent agreement with the exact results. Finally, we demonstrate that describing the slowdown of the boundary spin at and near the freezing frequency requires the incorporation of additional non-local terms. 

We initialize our system to a simple product state with all spins pointing along the $-x$ axis,
\begin{equation}
    |\psi_0 \rangle = |{-X} \rangle \equiv \bigotimes_{i=1}^{N} |{\leftarrow}\rangle_i.
\end{equation}
In the absence of an applied magnetic field ($h=0$), this state, being the maximally polarized ferromagnet along the $-x$ direction, is an exact eigenstate of $H_0$ with energy $E=0$ for odd $N$ and $E=1/4$ for even $N$. When decomposed in terms of $S^{z}_\text{tot}$ 
projected eigenstates, we get
\begin{equation}
\ket{\psi_0} = \frac{1}{2^{N/2}} \sum_{r=0}^N \sqrt{^NC_r}\ket{S^z_\text{tot}=\frac{2r-N}{2}}.
\end{equation}
%All states in R.H.S.\@ and consequently also $\ket{\psi_0}$ are eigenstates of $H_0$ in absence of the magnetic field $(h=0)$, with energy $E=0$ for odd $N$ and $E=1/4$ for even $N$. 
When a nonzero z-field ($ h \neq 0$) is turned on, the projected $S^{z}_\text{tot}$ states (on the R.H.S.) remain eigenstates of $H_0$, but their degeneracy is split. Thus $\ket{\psi_0}$ no longer remains an eigenstate of $H_0$. These projected states were shown to have low entanglement~\cite{Lee_PRBR2020}, in violation of the expected volume law at infinite temperature. Thus, this state serves as perhaps the simplest example of an embedded superspin (at high energy) in an otherwise chaotic spectrum, this is the basic idea in the core of perfect QMBS~\cite{Lee_PRBR2020, tunneltower}.

%$\ket{\psi_0}$ is an infinite temperature state w.r.t.\@ $H_0$ since its energy expectation value $\bra{\psi_0}H_0\ket{\psi_0}=0$ ($1/4$) resides at (near) the middle of the spectrum for odd (even) $N$.
Though the system initialized with $|{-X}\rangle$ 
exhibits perfect oscillations (in the Loschmidt echo and various observables) under the action of $H_0$, the situation is starkly different when the system is subject to a local periodic drive [Eq.~\eqref{eq:protocol}].
In previous work~\cite{melendrez2022localdrive}, we found that there are at least two parametric regimes depending on the kick strength and frequency. Typically, the driven spin thermalizes collectively with the rest, however, when a freezing condition involving the drive strength and frequency is satisfied, the driven spin is almost (but not strongly) decoupled from the rest of the system. We will elaborate on this in the next subsection.

In what follows we have worked primarily with the $|{-X}\rangle$ initial state, however, we emphasize that all our results remain insensitive to this choice and are valid for any product state in $S^x$-basis.

\subsection{Failure of the zeroth order Floquet Hamiltonian}

We perturbatively construct and analyze the Floquet Hamiltonian, $H_F$, using a resummed F-M expansion (as charted out in detail in Appendix~\ref{app:A}). In this scheme, $H_F=\sum_{l=0}^{\infty}H_F^{(l)}$
where the various orders are denoted by $H_F^{(l)}$ and, in principle, can be calculated for each $l$. 

A computation of the $l=0$ term gives, 
\begin{widetext}
\begin{align}
    H_F^{(0)}&=-S^x_1S^x_2-\frac{\sin\lambda}{\lambda}(S^y_1S^y_2+S^z_1S^z_2)+\frac{1-\cos\lambda}{\lambda} (S^z_1S^y_2-S^y_1S^z_2)-\frac{h}{\lambda}[\sin\lambda S^z_1+(1-\cos\lambda)S^y_1]\nonumber\\
        &\quad+\sum_{i=2}^{N-1}(-1)^{i}{\bf S}_i\cdot{\bf S}_{i+1}-h\sum_{i=2}^{N}S^z_i,
\end{align}
\end{widetext}
where $\lambda \equiv {\gamma \tau}/{2} = {\pi \gamma}/ {\omega} $. 
Additionally, the calculation for $l=1$ shows that $H_F^{(1)}$ vanishes for all drive parameters. 

When $\lambda = 2 \pi k$, where $k$ is an integer, the only non-zero term involving the driven boundary site in $H_F$ (calculated up to $l=1$) is $-S_1^{x} S_2^{x}$. This suggests that the edge spin can be \textit{completely} frozen if one tunes the drive frequency to $\omega_c^k = \gamma / (2k)$ since the following commutation relation
\begin{equation}
[S^x_1,H_F^{(0)}+H_F^{(1)}]=0
\end{equation}
is exactly satisfied.

\begin{figure}[t!]
    \centering
    \includegraphics[width=\linewidth]{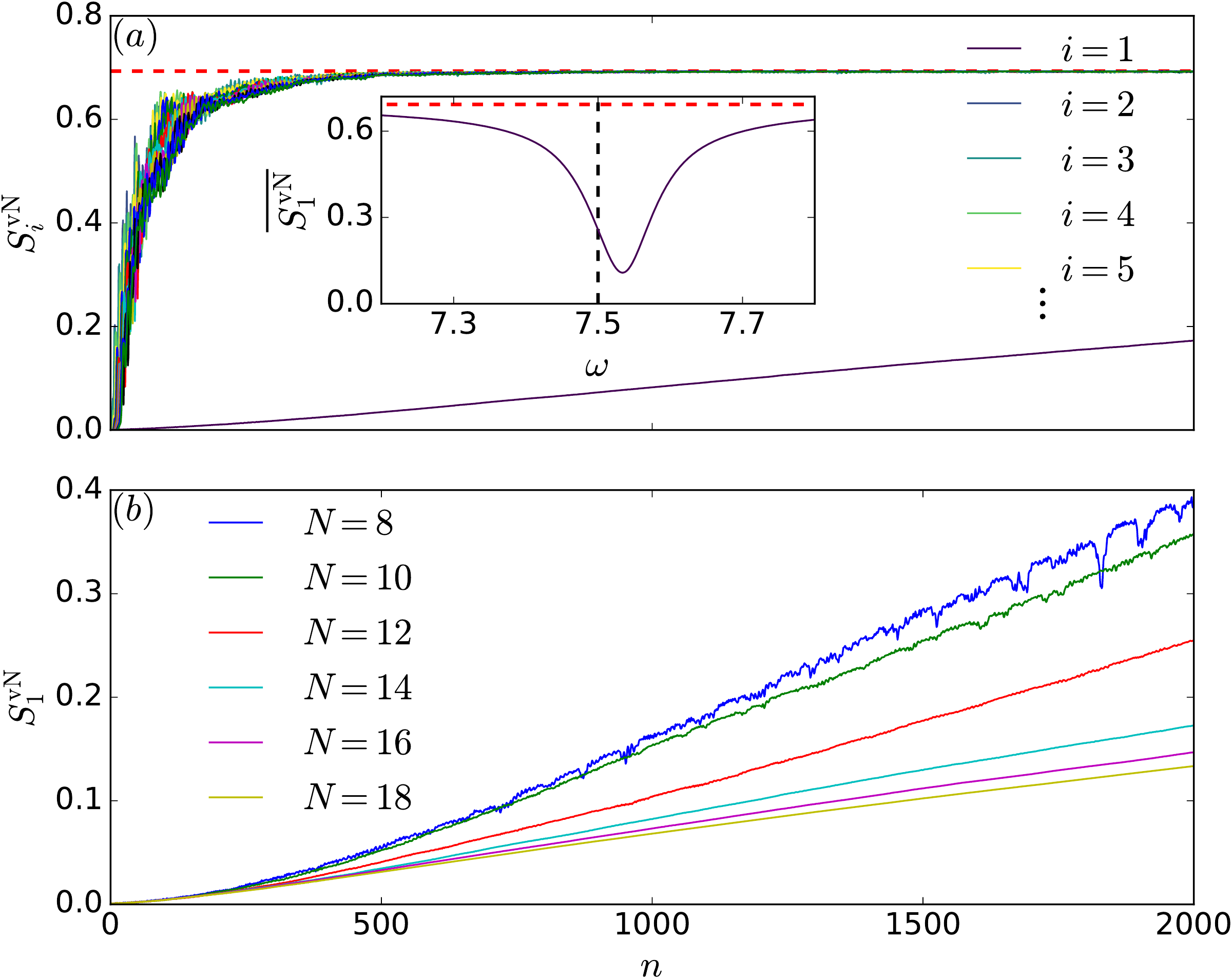}
    \caption{Stroboscopic dynamics of (a) single site entanglement entropy $S^{\text{vN}}_i$ for all the sites ($i$) for $N=14$. The red (dashed) line is used to denote the maximum single-site entropy ($\ln 2$). $\overline{S^{\text{vN}}_1}$ (averaged $S^{\text{vN}}_1$ over first 2000 cycles) is plotted with $\omega$ in the inset showing the absence of dynamic freezing of the edge spin even for the most appropriate choice of the drive frequency. A vertical dashed line is also drawn at $\omega^1_c=7.5$ to clearly show the minima in $\overline{S^{\text{vN}}_1}$ ($\omega^1_m$) is shifted from $\omega^1_c$. (b) entanglement entropy of the boundary site ($S^{\text{vN}}_1$) for different system sizes ($N$).  $h=1$, $\gamma=15$, $\omega=\omega^1_m=7.53$ for both the panels. The $N=16$ and $N=18$ calculations were done with matrix-product-state-based TEBD calculations for a time step of $\delta t = \tau/50$ and a maximum bound dimension of $2^{N/2}$.}
    \label{fig:site1}
\end{figure}

We focus on the $k=1$ case, though all the discussions remain valid for any $k$. Fig.~\ref{fig:site1} (a) presents the stroboscopic dynamics of $S^{\text{vN}}_i$, for all $i$ with $s_d=\{1\}$. We find that the relaxation rate of the boundary spin is indeed very small  at $\omega\sim\omega^k_c$ (compared to other sites which thermalize very quickly in spite of being undriven), however, it is always nonzero. This can be seen in the inset of Fig.~\ref{fig:site1}(a), where we plot $S^{\text{vN}}_1$ averaged over the first 2000 cycles with $\omega$ near $\omega^1_c$. The frequency which gives the slowest relaxation ($\omega=\omega^k_m$) shifts to a somewhat higher value compared to $\omega^k_c$ (i.e.\@ $\omega^k_m>\omega^k_c$). This result can not be explained by the lowest-order ($l=0$) Floquet Hamiltonian.

We also monitor the size dependence of the relaxation rate of $S^{\text{vN}}_1$ in Fig.~\ref{fig:site1}(b) using a combination of exact diagonalization for $N \leq 14$, 
and matrix product state based time-evolving block decimation (TEBD) technique~\cite{vidal2004,white2004} for $N>14$, setting the bond dimension to its maximal value of $2^{N/2}$. A time step of $\delta t = \tau/50$ was used for the TEBD calculations\cite{itensor}, and we checked there was almost no difference in our results for $\delta t \sim \tau/30$ to $\delta t \sim \tau/100$. 
While increasing the system size, we find that the relaxation of the edge spin slows down, but most likely not exponentially. Moreover, the relaxation dynamics tends to be saturated beyond a threshold system size ($\approx 16$ as can  be seen in Fig.\ref{fig:site1}(b)), an observation that is consistent with the phenomenology of an approximate SZM \cite{kemp}.

\subsection{Higher order corrections and freezing condition} 

To explain our numerical observations, we need to go to higher terms in the F-M expansion. However, before we dive into this calculation, we note that the F-M expansion to obtain the Floquet Hamiltonian for interacting systems is plagued by various convergence issues~\cite{kuwahara}.
Though these issues can be circumvented to some extent by systematic resummations in certain parameters (e.g. drive frequency) ~\cite{replica,anatoli1,anatoli2}, the calculation of the resummed expansion becomes increasingly difficult in higher order in other parameters (e.g. drive amplitude) due to the presence of nested commutators of many-body terms. In spite of this, the calculation of higher-order terms in F-M has been fruitful, as it can reveal physical effects not accessible in the lower order expansion.

As anticipated, the calculation of $H_F^{(2)}$ is somewhat complicated by the existence of multiple terms, for completeness the various contributions have been described in Appendix~\ref{app:A}. Here we categorize the resulting expression into ``local'' and ``non-local'' parts, 
\begin{equation}
H_F^{(2)}=H^{(2)}_{F,\ \text{loc}} + H^{(2)}_{F,\ \text{nonloc}}.
\end{equation}
The ``local'' part, consisting of only one- and two-body (n.n.\@) operators (including the boundary site), can be written as 
\begin{widetext}
\begin{equation}
H^{(2)}_{F,\ \text{loc}} = a_1(\boldsymbol{p})S^z_1 + a_2(\boldsymbol{p})S^y_1 + a_3(\boldsymbol{p}) S^z_1 S^z_2 + a_4(\boldsymbol{p})S^y_1S^y_2 + a_5(\boldsymbol{p})S^z_1S^y_2 + a_6(\boldsymbol{p})S^y_1S^z_2 + \cdots,
\end{equation}
\end{widetext}
where $\boldsymbol{p}\equiv(h,\gamma,\tau)$ represents the set of drive parameters (see Appendix~\ref{app:A} for the full expression of all the $a_i$ coefficients). 
The ellipsis designates terms, which commute with $S^x_1$. 
It is important to note that our classification of ``local'' is based on the fact that the terms have the same functional forms as those appearing in $H_F^{(0)}$, hence only renormalizing their strength. 

Interestingly, we find that $a_i$s contain terms with similar magnitude as the zeroth ($l=0$) term [i.e.\@ $\mathcal{O}(1/\gamma)$], some of which can be nonzero even at the special freezing frequencies $\omega_c^k$. Collecting only the $\mathcal{O}(1/\gamma)$ terms from $H^{(2)}_{F,\ \text{loc}}$, we rewrite $H_F$ as:
\begin{widetext}
\begin{eqnarray}
  && \hskip -0.5cm H_{F,\text{loc}}[\mathcal{O}(\frac{1}{\gamma})]\nonumber\\
  &=& H_F^{(0)} + H_F^{(1)} + H^{(2)}_{F,\ \text{loc}}\left[\mathcal{O}\left({1}/{\gamma}\right)\right]\nonumber\\
      &=& -S^x_1S^x_2 - \frac{2\sin\lambda}{\gamma \tau}
      \left[
        \left(1+\frac{(1+4h^2) \tau^2}{48}\right)S^y_1S^y_2 + \left(1+\frac{ \tau^2}{48}\right)S^z_1S^z_2
      \right] + \frac{2}{\gamma \tau}
      \left[
        \left(
          1-\cos\lambda-\frac{(1+4h^2) \tau^2(2+\cos\lambda)}{48}
        \right)S^z_1S^y_2
      \right.
  \nonumber\\
  &&\quad-
    \left.
      \left(1-\cos\lambda-\frac{\tau^2(2+\cos\lambda)}{48}\right)S^y_1S^z_2
    \right] - \frac{2h}{\gamma \tau}
    \left[
      \sin\lambda \left(1+ \frac{\tau^2}{48}\right) S^z_1
      + \left(1-\cos\lambda - \frac{\tau^2(2+\cos\lambda)}{48}\right)S^y_1
    \right]
  \nonumber\\
  &&\quad+\sum_{i=2}^{N-1}(-1)^{i}{\bf S}_i\cdot{\bf S}_{i+1}-h\sum_{i=2}^{N}S^z_i.
\label{eq:HF2_1overgamma}
\end{eqnarray}
\end{widetext}
With this expression it becomes clear that $[S^x_1,H_F] \ne 0$ at $\omega = \omega_c^k$. In fact, there is not a single ${\boldsymbol p}$ for which all $S^x_1$ non-commuting terms in Eq.~\eqref{eq:HF2_1overgamma} simultaneously go to zero. In other words, it is impossible to dynamically freeze the edge spin via the emergence of a corresponding local conserved quantity. 
This result must be contrasted with the case of $H_F^{(2)}$ for the case of global driving with a square pulse~\cite{asmi}, where the contributions are expected to be at least $\mathcal{O}(1/\gamma^2)$ or smaller (see Appendix~\ref{app:globaldriving}). 

\begin{figure}
    \centering
    \includegraphics[width=\linewidth]{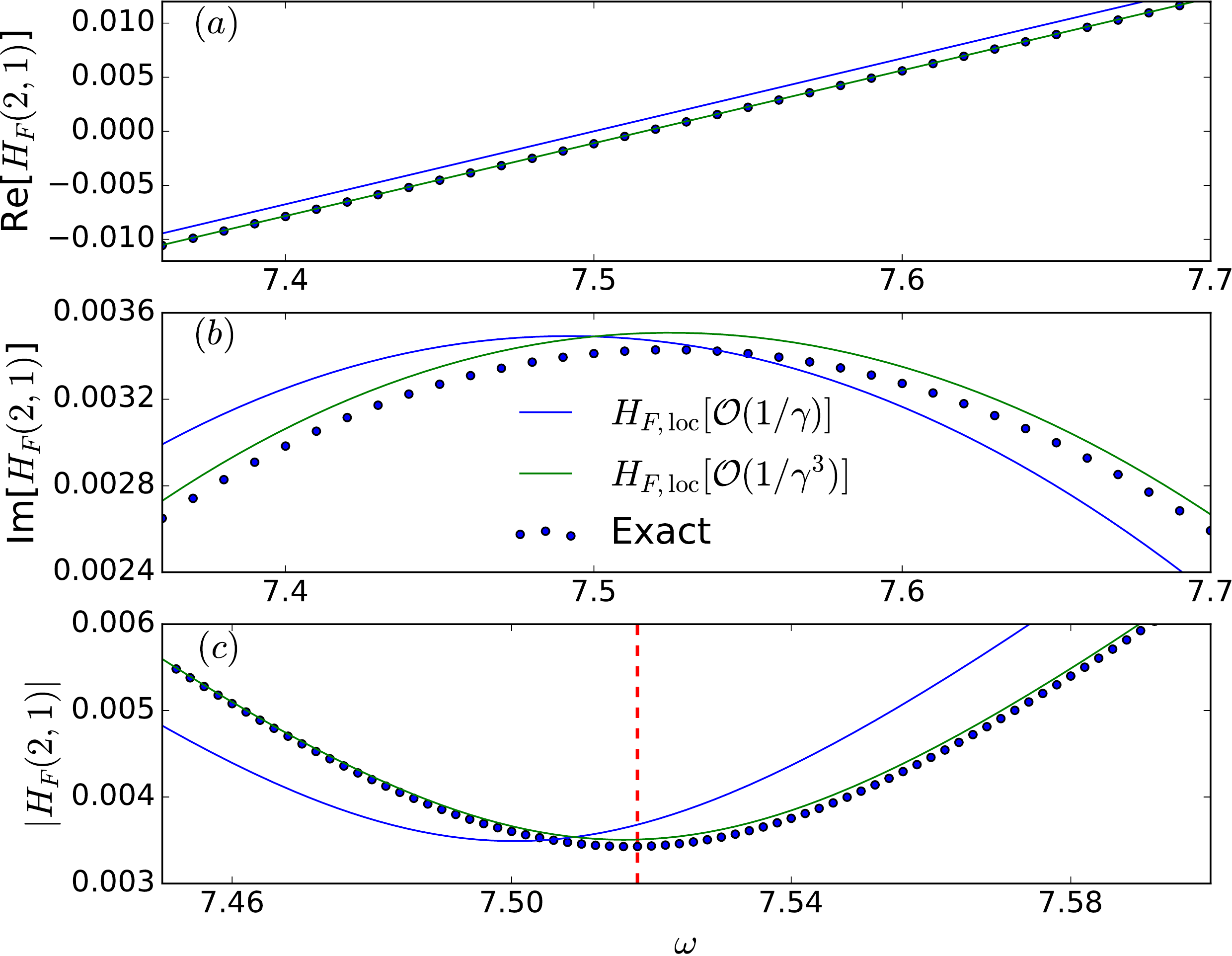}
    \caption{Comparison of (a)~real, (b)~imaginary, and (c)~absolute value of $H_F(2,1)$ between exact numerics (black dots) and expressions obtained from F-M expansions. Blue (green) curve is obtained from Eq.~\ref{eq:HF2_1overgamma} (Eq.\ref{eq:HF2_1overgamma3}). A vertical red (dashed) line is used in panel (c) to guide the eye to the minima of $H_F(2,1)$ which is clearly shifted from $\omega^1_c=7.5$. $h=1$, $\gamma=15$, $N=8$. } 
    \label{fig:matching}
\end{figure}

The local terms, to order $1/\gamma$, do not significantly affect the freezing condition obtained from $H_F^{(0)}$, they alone are quantitatively inadequate for explaining the observed shift in the freezing frequency. Thus we retain higher order local terms in $H_{F,\text{loc}}^{(2)}$ and define
\begin{equation}
    H_{F,\text{loc}}\left[\mathcal{O}\left(\frac{1}{\gamma^3}\right)\right] \equiv H_F^{(0)}+H_F^{(1)}+H_{F,\text{loc}}^{(2)}\left[\mathcal{O}\left(\frac{1}{\gamma^3}\right)\right].
    \label{eq:HF2_1overgamma3}
\end{equation}
To demonstrate that the Hamiltonian in Eq.~\ref{eq:HF2_1overgamma3} captures the shift of the point of minimum relaxation to a frequency slightly higher than $\omega^k_c$, we calculate a matrix element of $H_F$ in the $S^x$-basis, $H_F(2,1)$ where $\ket{1}=\ket{\psi_0}$ and $\ket{2} = \sigma^z_1 \ket{1}$.
%, using F-M expansion and (see Appendix~\ref{app:A} for the consolidated expressions) 
Fig.~\ref{fig:matching} shows a comparison of different levels of approximation for $H_F(2,1)$ with the exact numerical calculation.
One can see that $|H_F(2,1)|$ obtained from Eq.~\eqref{eq:HF2_1overgamma}, though always nonzero (supporting the absence of freezing of the edge spin), does not match the exact results, particularly it still shows the minimum at a frequency which is extremely close to $\omega^k_c$, rather than the shifted value. In comparison, including the local $\mathcal{O}(1/\gamma^2)$ and $\mathcal{O}(1/\gamma^3)$ terms in $H_F^{(2)}$ yields substantially good agreement with the exact numerics, in particular the location of the minima of matrixelements (like $H_F(2,1)$) which determines the freezing frequency.

Going beyond individual matrix elements, we compare the exact 
stroboscopic dynamics with that generated by the $H_{F,\text{loc}}[\mathcal{O}(1/\gamma^3)]$ in Fig.~\ref{fig:matching_dynamics}. 
This approximation is excellent for capturing the dynamics of all spins away from the freezing frequency. At and near the freezing frequency, it captures the dynamics of all, but the boundary spin, accurately. Thus the local $H_F$ approximation can capture not only the individual matrix elements but also the stroboscopic dynamics quite accurately at least for short times. 

\subsection{Role of non-local terms to slowdown the edge-spin}
\label{non-local}
\begin{figure}
    \centering
    \includegraphics[width=\linewidth]{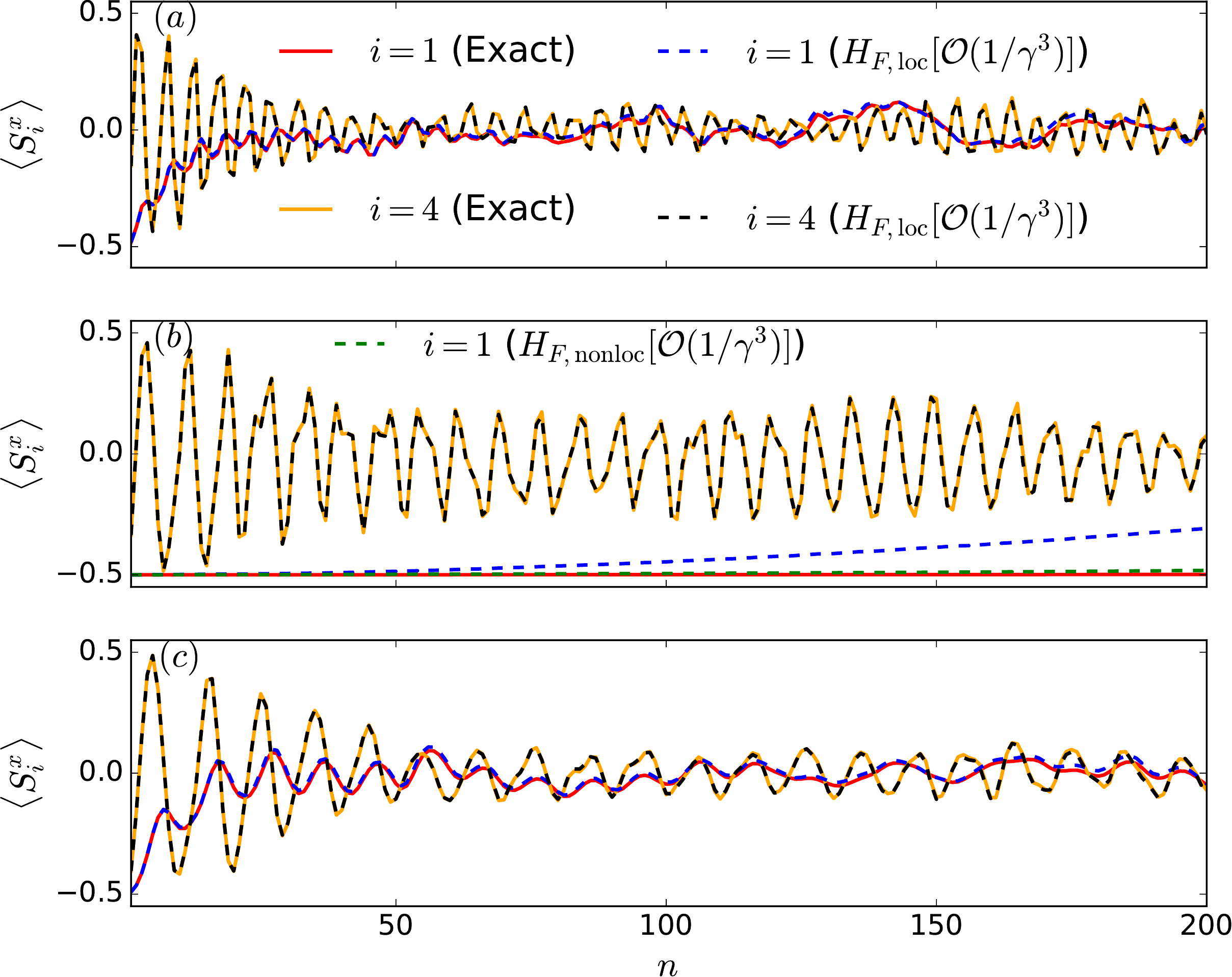}
    \caption{Comparison of stroboscopic dynamics from exact numerics and analytical $H_F$ obtained from the resummed F-M expansion at (a) $\omega=5$ (b) $\omega=7.53$ (c) $\omega=10$. $\gamma=15$, $h=1$, $N=8$. Away from the freezing frequency the stroboscopic dynamics is well captured for all spins within a ``local'' approximation for $H_F$, in comparison to exact numerics. At or near the freezing frequency, non-local terms are required to accurately capture the behavior of the driven boundary spin.} 
    \label{fig:matching_dynamics}
\end{figure}

Interestingly, the failure of the local $H_F^{(2)}$ only for the dynamics of the boundary spin at and near $\omega^k_m$ reveals the importance of the non-local terms, that we have ignored up to this point. We have numerically checked that near the freezing frequency many matrix elements of $H_F$, generated by long-range and multi-spin interaction terms involving the boundary spin are quite significant relative to the local matrix elements [like $H_F(2,1)$ which shows a minimum at this parameter regime]. We find that such non-local terms, appearing in $H_F^{(2)}$, can at most be four-body operators having support on sites 1--4. We have calculated all such terms (dubbed as $H^{(2)}_{F,\text{nonloc}}$) with their respective strengths in terms of the drive parameters (see Appendix~\ref{app:A}). Such terms when included in $H_F$ are found to improve the agreement with exact numerics for the boundary spin (see Fig.~\ref{fig:matching_dynamics}(b)). 

The non-local terms, though very weak, will proliferate in further higher orders; a careful inclusion of them may increase the timescale of the agreement with exact numerics. In fact we find that the non-local and multipsin terms already present in $H^{(2)}_{F,\text{nonloc}}$ are sufficient to capture the slow dynamics of the edge spin quite accurately up to $n \sim 200$. The strength of the nonlocal terms calculated in 2nd order ($l=2$)  will be renormalized in higher order in the F-M expansion which in turn will extend the agreement with exact dynamics to longer times. This is quite nontrivial for two reasons. First, any nonthermal (or prethermal) phenomenon in a Floquet system is naively expected to be linked with the existence of a local $H_F$ whereas a non-local $H_F$ usually promotes heating. But here we observe the exact opposite, the thermalizing dynamics of the bulk sites is well captured by a local $H_F$ (even at $\omega^k_m$ as shown in Fig.~\ref{fig:matching_dynamics}) but not the freezing dynamics of the boundary site which necessitates the presence of certain long-range terms for its onset. Here we also note that $H^{(2)}_{F,\text{nonloc}}$ is completely off-diagonal in the 
$x$-basis and the proliferation of these terms is expected to dephase the initial state $\ket{\psi_0}$.
Second, from a practical point of view, 
it establishes the freezing phenomenon reported in this work as a genuine drive-induced many-body effect, which is challenging to engineer with fine-tuned long-range terms in a time-independent Hamiltonian.

As noted previously, the rate of relaxation of the boundary spin at the special frequencies ($\omega^k_m$) decreases further with increasing system size, as can be seen in Fig.~\ref{fig:site1} (b).
This is at odds with the usual expectation that increasing the system size generically increases the phase space available for thermalization and should thus speed up the process.
%generically ergodicity gets more and more uncovered as we increase the system size. 
This apparent anomalous behavior already suggests that boundary driving alone can be enough to trigger the onset of an approximate SZM \cite{kemp,aditi1} in the system. However, %\hjc{unclear scientific idea, exponential in what?}, 
the relaxation time of the boundary spin is significantly shorter than the time associated with a SZM, in which case it is exponentially large in system size.

The inability of boundary-site driving to fully freeze the boundary spin and the presence of non-local terms motivates the next natural question: what happens when multiple sites (including the boundary one) are driven together? Is there any requirement on the number of driven sites as well as the specific locations where the drive should be applied to maximize the lifetime of the edge spin? We address these questions in the next section. 

\section{Emergence of the strong zero mode via multi-site driving}
\label{sec:multisite}
%\begin{itemize}
%\item Numerical experiments demonstrating the optimal driving sites\\
%\item TEBD
%\item just motivate SZM here\\
%\end{itemize}

\subsection{Optimal drive protocol}

We now consider a situation when multiple spins in the bulk are driven simultaneously, in addition to the boundary. Intuitively, the increase in the number of driven sites would lead to more rapid thermalization of all the spins. As discussed in Sec.~\ref{non-local}, each driven site generates a larger collection of non-local terms which slow down the edge spin near the freezing frequency. In this section we first explore the landscape of few-site protocols and investigate their effects on the edge spin. This subsequently leads us to the development of a multi-site driving protocol, resulting in the further slowdown of the edge spin.

\begin{figure}
    \centering
    \includegraphics[width=\linewidth]{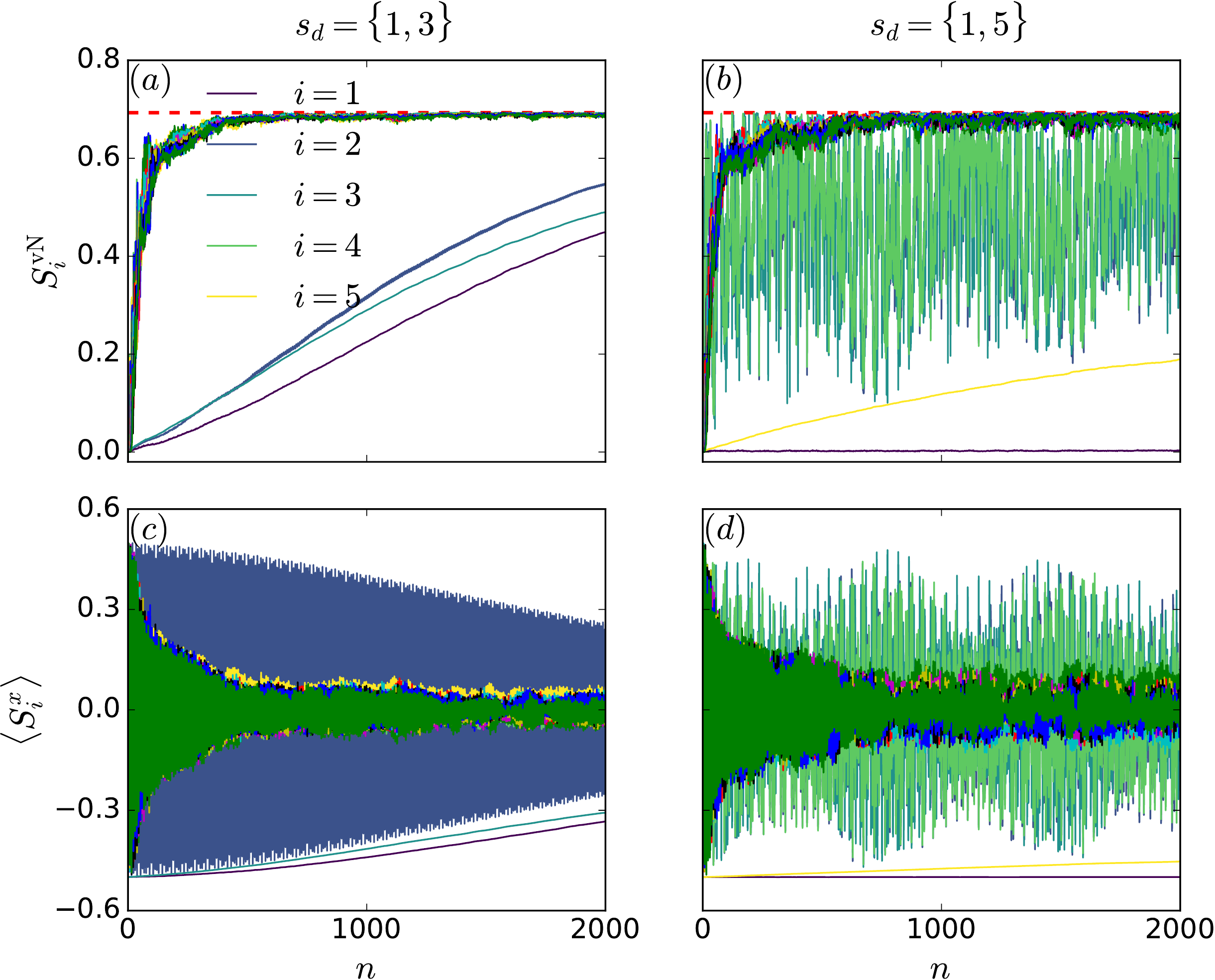}
    \caption{Stroboscopic $S^{\text{vN}}_i$ (upper panels) and $\langle S^x_i \rangle$ (lower panels) for different two-site driving. Locations of the driven sites ($s_d$) are mentioned at the top of each column. $h=1,\gamma=15,\omega=\omega^1_m=7.53,N=14$. }
    \label{fig:twosite}
\end{figure}

When the boundary and a generic bulk site are driven together, our results indicate that the slowdown of the boundary site is not improved. We demonstrate this behavior with an example for $s_d=\{1,3\}$ in Fig.~\ref{fig:twosite}(a,c), where we observe that the relaxation of the boundary site is faster than the case of $s_d=\{1\}$. The undriven sites are found to thermalize quickly, with the exception of the intermediate spins, which exhibit slower relaxation to thermal equilibrium. However, remarkably we find that a particular two-site driving protocol results in strong freezing of the edge spin: when in addition to the boundary site the fifth site is also driven ($s_d=\{1,5\}$). Indeed, the first site shows almost no relaxation of $S_1^x$ for the drive cycles plotted in Fig.~\ref{fig:twosite}(b,d). The intermediate sites assume a different character now, showing a strong temporal oscillation before eventually decaying out to zero. This particular driving protocol also causes a widened frequency window which admits slow (but nonzero) relaxation of the edge spin. Moreover, we find that the relaxation of the edge spin, analogous to the boundary-driven case, slows down with increasing the system size [see the inset in Fig.~\ref{fig:dynamics2}].

This freezing behavior is necessarily a high-order many-body effect, as the drive itself only produces interactions with a range of three sites away from the driven one (in the 2nd term in F-M). Therefore, driving site 5 does not change the structure of the effective Hamiltonian from Eq.~\eqref{eq:HF2_1overgamma}, only adding terms supported on sites 2--8. This implies that the origin of the suppression of the boundary relaxation is a higher-order many-body process, where the driven site 5 impacts sites 2--4, which in turn impact the boundary. This complex collective behavior seems to be a unique property of only this particular two-site protocol ($s_d=\{1,5\}$), and will become the key ingredient in our subsequent proposal for the Floquet engineering of the SZM.

We also note that the additional driven site is itself not frozen. This can be understood by noticing that the Hamiltonian now includes newly generated three-body interaction terms (e.g.\@ $S^x_4 S^z_5 S^x_6$, see Appendix~\ref{app:B}), which renormalizes the field strength at site 5. Note that such a term cannot occur for the boundary-driven case. To illustrate this one can, for example, compare the aforementioned matrix element $H_F(2,1)$ with the matrix element representing single spin-flip events at site 5, $\bra{\psi_0} H_F \sigma^z_5 \ket{\psi_0}$. Indeed, we find the latter to be larger in magnitude, contributing to a much faster dephasing of site 5 as compared to site 1.

\subsection{ 
Freezing of edge spin on increasing the number of driven sites}
Motivated by our observation of the slowdown of the edge due to the two-site protocol, we consider the driving of additional sites. Increasing the number of driven sites further, we find that generically there is a rapid relaxation of all sites, including the boundary. However, when the driving sequence is specifically chosen as $s_d = \{1, 5, 9\}$, the freezing of the boundary site is significantly enhanced compared to the two-site protocol with $s_d=\{1, 5\}$. We can conclude that the further slowdown has a similar origin to the two-site protocol: a high-order many-body process, this time involving site 9. Based on the spatial periodicity of the few-site protocols, we propose the following Floquet engineering protocol to further slowdown the edge spin, where the driving sequence is composed of the cluster $s_d = \{1, 5, 9, ..., 4k+1\}$. Subsequently, we will use the term ``every fourth site'' to describe this cluster, and designate the total number of driven sites as $n_d^\text{4th}$. The existence of this protocol does not preclude other protocols which can stabilize a SZM, and we consider this as a particular case in a general class of locally driven protocols.

\begin{figure}
    \centering
    \includegraphics[width=\linewidth]{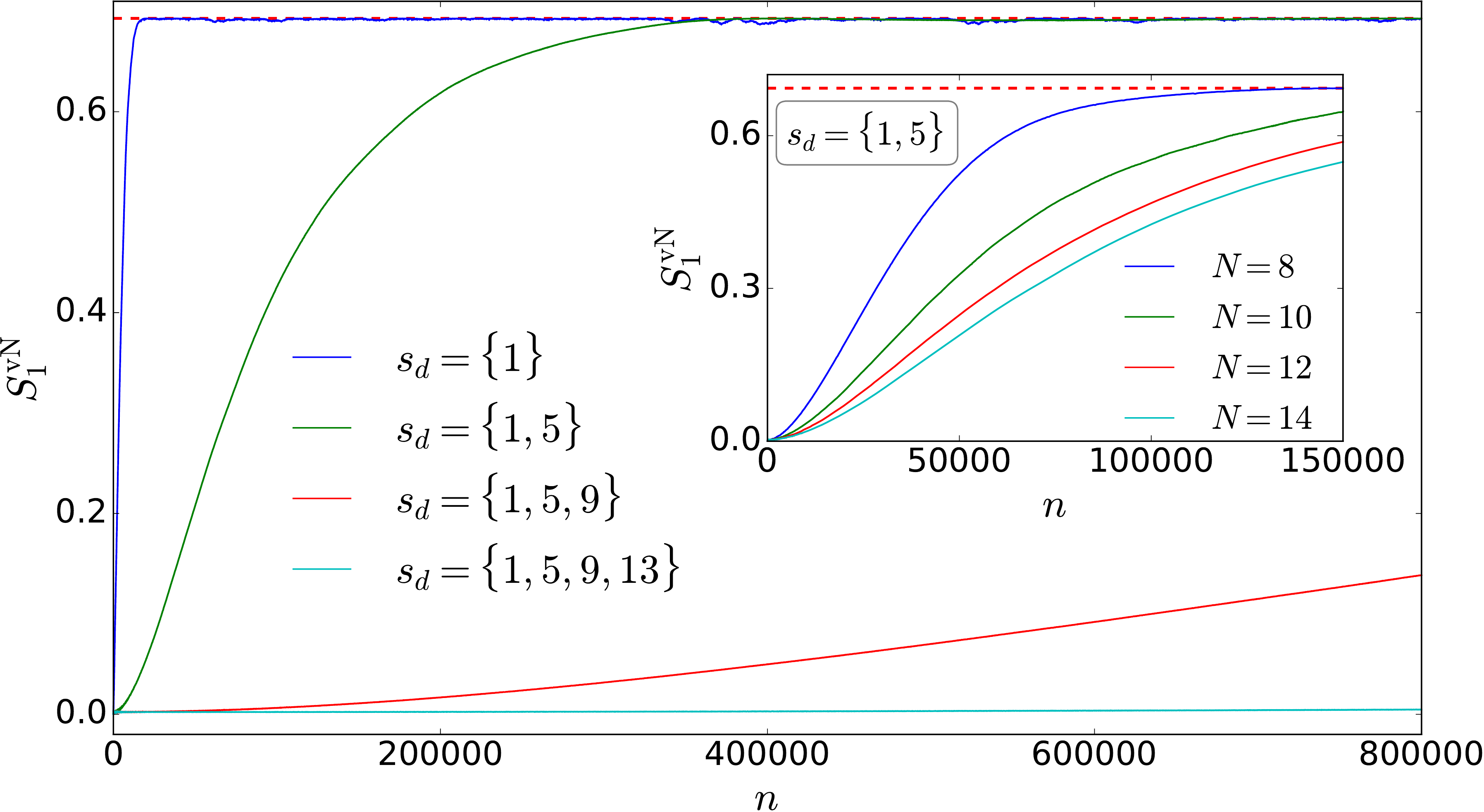}
    \caption{Long time stroboscopic dynamics of $S^{\text{vN}}_1$ for different multi-site driving showing the optimal choice of drive locations (every fourth site) to freeze the boundary spin. Inset shows $S^{\text{vN}}_1$ vs $n$ for $s_d=\{1,5\}$ for different system sizes. All parameters are the same as in Fig.~\ref{fig:twosite}. }
    \label{fig:dynamics2}
\end{figure}

In Fig.~\ref{fig:dynamics2} we show how the growth of entanglement entropy of the boundary site decreases while increasing the size of the cluster $s_d$: we find a hierarchy of freezing behaviors with the growing number of driven spins. This conclusion is supported by the exact diagonalization results allowing us to reach large number of drive cycles, see Fig.~\ref{fig:dynamics2}. Furthermore, the inset in Fig.~\ref{fig:dynamics2} shows that although increasing the system size alone improves the freezing behavior, the slowdown can be most effectively achieved by also increasing the number of driven sites.

Intuitively, the extremely slow dynamics in the proposed protocol is directly related to the gaps between quasi-energy levels in the Floquet spectrum. This can be seen by considering thermalization of the expectation value of a local observable for a driven state $\ket{\psi(n\tau)}$~\cite{Deutsch1991, Srednicki1994, Rigol2008}: 
\begin{align}
\langle{\psi(n\tau)}| S^x_1 |{\psi(n\tau)}\rangle & = \sum_{q \ne r} c_r^* c_q e^{i(\epsilon_r-\epsilon_q)n \tau}\langle{\Phi_r}| S^x_1 |{\Phi_q}\rangle \\
& \quad + \sum_r |c_r|^2 \langle{\Phi_r}| S^x_1 |{\Phi_r}\rangle, \nonumber
\end{align}
where $\epsilon_r$ and $\ket{\Phi_r}$ are the $r$-th quasienergy and eigenstate in the first Floquet Brillouin zone (BZ). The first sum in the equation controls the relaxation speed, while the second sum gives the long-time expectation value in the thermalized state (diagonal ensemble average). %For infinite temperature states the diagonal ensemble average tends to zero, therefore the \hjc{check short time fit $log (1 + t/tau^{*})$} exponentially slow dynamics of $S_1^x$ is attributed to the characteristics of the off-diagonal element. 
For our protocol, the zeroth-order effective Floquet Hamiltonian $H_F^{(0)}$ along with the $2^{\text{nd}}$ order corrections, are invariant under $S^x \rightarrow -S^x$ transformation, which constraint the diagonal ensemble average of $S_1^x$ to be zero. Since our exact calculations of the expectation value of $S_1^x$ in Floquet eigenstates also yield zero up to machine precision for a wide range of parameters, the invariance is expected to exist at all orders of $H_F$.  

For generic non-integrable systems the off-diagonal contribution decays to zero quite rapidly. But here we find that the energy gaps between the states (say $r$ and $q$) which dominate the matrix elements are extremely small. For those states  $|\epsilon_r - \epsilon_q|$ ranges from $\sim 10^{-5}$ for $s_d = \{1,5\}$ dropping to $\sim 10^{-8}$ for $s_d = \{1,5,9,13\}$. This is further supported by the level statistics of the Floquet Hamiltonian, which shows a strong Poisson distribution and the absence of level repulsion, putatively indicating the emergence of a symmetry or fragmentation of the Hilbert space (see Appendix~\ref{app:levelstat}). We provide a coherent analysis of this extensive quasi-degeneracy throughout the entire Floquet spectrum, which shows that the freezing ultimately has its origin in the emergence of the SZM.

\subsection{Emergence of the strong zero mode}
\label{sec:szm}

In this section, we elucidate the step-by-step emergence of the strong zero mode which is the crux of the exponentially slow dynamics of the edge mode operator. We expect that a system endowed with a SZM, as defined in Sec.~\ref{sec:model}c, is characterized by a spectrum-wide (quasi)-degeneracy and relaxation of the edge that can be slowed down by increasing the system size. 

\begin{figure}
 \includegraphics[width=\linewidth]{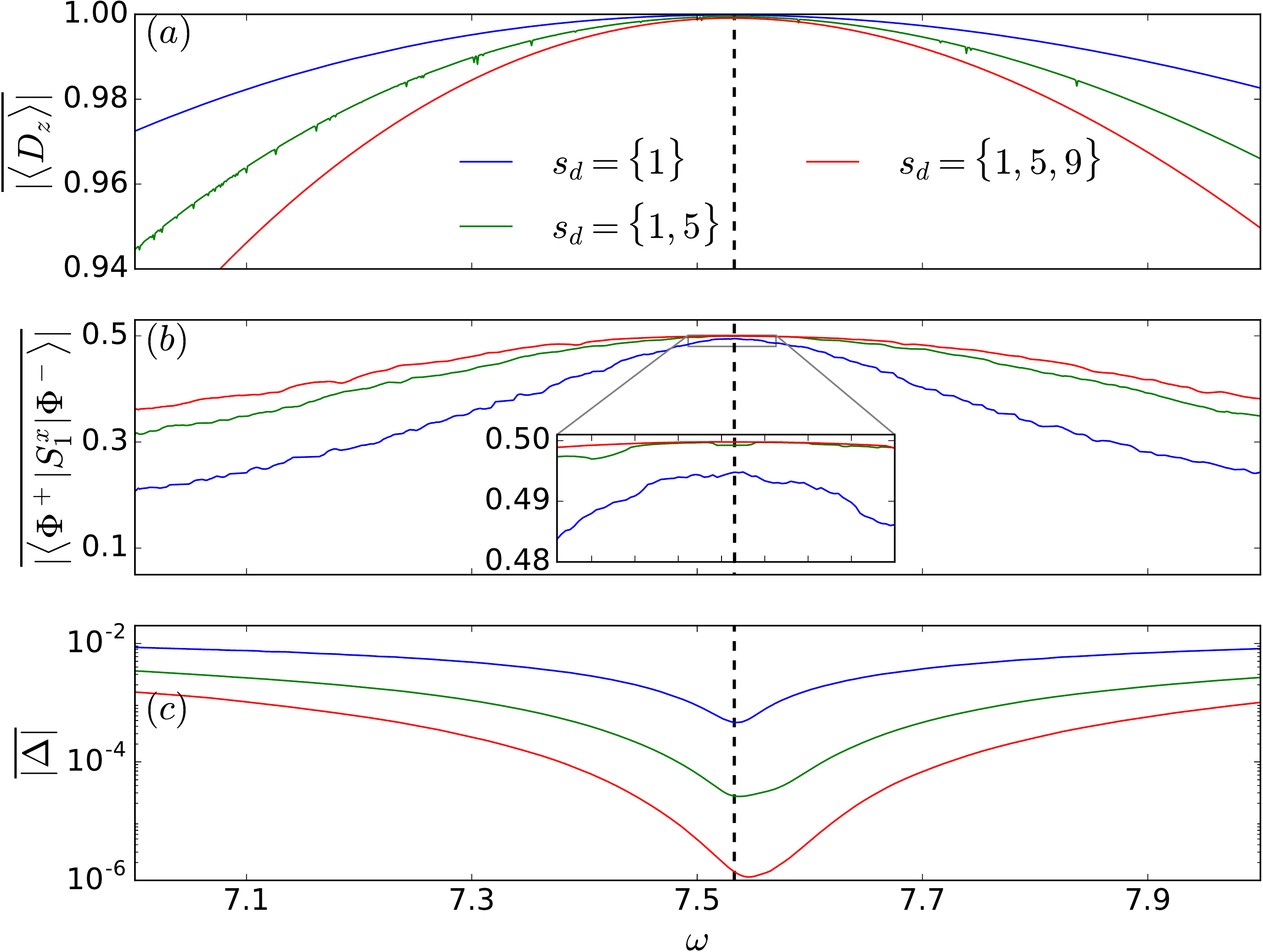}
 \caption{Emergence of strong zero-mode:
 (a)~$\overline{|\langle D_z\rangle|}$ vs $\omega$ showing the emergence of  a discrete symmetry, $D_z=\prod_i(2S^z_i)$, at $\omega\sim \omega^1_m$.
 (b)~$\overline{|\langle \Phi^+|S^x_1|\Phi^-\rangle|}$ vs $\omega$. Inset shows that the condition $S^x_1|\Phi^{\pm}\rangle=\frac{1}{2}|\Phi^{\mp}\rangle$ is satisfied better for $s_d=\{1,5\}$ and $s_d=\{1,5,9\}$ compared to $s_d=\{1\}$.
 (c)~$\overline{|\Delta|}$ vs $\omega$ showing a deep at $\omega\sim\omega^1_m$ which implies the emergence of SZM induced quasi-degeneracy between the Floquet eigenvalues from different $D_z$ sectors. $N=10$, $\gamma=15$ for all the plots. We have performed a moving average over the raw data (in an $\omega$ interval of 0.008) for panel (b). The vertical dashed line is a guide to the eye to the position of strongest freezing.} 
\label{fig:szm}
\end{figure}

First, let us consider the existence of a discrete symmetry operator. The undriven system has a $\mathbb{Z}_2$ symmetry of $D_z = \prod_i \sigma^z_i = \prod_i (2 S^z_i)$, yet the drive in the $x$-direction generally destroys it. However, we can exploit the freedom in the driving parameters and note that near the special frequencies $\omega^k_m$ the terms in Eq.~\eqref{eq:HF2_1overgamma} that do not commute with $D_z$ vanish. We test this for the full Floquet Hamiltonian numerically in Fig.~\ref{fig:szm}(a), where we plot $\overline{|\langle D_z \rangle|} = \sum_{r=1}^{2^N} |\langle\Phi^r| D_z |\Phi^r\rangle| / 2^N$; this quantity approaches 1 when each Floquet eigenstate is simultaneously an eigenstate of the proposed symmetry operator. We find that although $D_z$ is not an exact symmetry, it is nearly conserved at the special freezing frequencies. The formation of SZM requires a special pairing structure of the Floquet eigenstates and quasi-degenerate eigenvalues, along with the global discrete symmetry $D_z$.

We now label the Floquet eigenstates by the eigenvalues of the symmetry operator $D_z$ as $H_F\ket{\Phi^{\pm}_r}=\epsilon^{\pm}_r\ket{\Phi^{\pm}_r}$. We show that the eigenstates occur in pairs in the Floquet spectrum where the operator which toggles between the states $|\Phi_r^+\rangle$ and $|\Phi_r^-\rangle$ is localized at the edge. This operator anticommutes with the global symmetry operator $D_z$.
We propose that an approximate operator, which satisfies these properties, is $O_\text{szm} \approx S^x_1$. This is shown by the numerical calculation of the matrix elements between $|\Phi_r^+\rangle$ and $|\Phi_r^-\rangle$. In Fig.~\ref{fig:szm}(b) the average matrix element over all pairs of eigenstates   
\begin{equation}
\overline{|\langle\Phi^-| S^x_1 |\Phi^+\rangle|} = \sum_{r=1}^{2^{N-1}}|\langle\Phi_r^-| S^x_1 |\Phi_r^+\rangle|/2^{N-1},
\end{equation}
is plotted as a function of $\omega$. This quantity measures how $S_1^x$ connects the two symmetry sectors, and should approach $1/2$ when the pairing is exact, when the matrix element is non-zero for a single pair of Floquet eigenstates. Indeed, we find that increasing the number of driven sites leads to a more accurate pairing of the states. We propose an ansatz for the Floquet eigenstates with the following structure:
\begin{equation}
  |\Phi^{\pm}_r\rangle = \frac{1}{\sqrt{2}} \left( |{\rightarrow}\rangle \otimes |\xi^{\pm}_r\rangle \pm \vphantom{\frac{}{}} |{\leftarrow}\rangle \otimes |\bar{\xi}^{\pm}_r\rangle \right),
  \label{eq:Floquet_eigenstates}
\end{equation}
where states $|\xi^\pm_r\rangle$ and $|\bar\xi^\pm_r\rangle$ obey conditions $\langle \xi^{\pm}_r | \xi^{\pm}_q \rangle + \langle \bar{\xi}^{\pm}_r | \bar{\xi}^{\pm}_q \rangle = 2 \delta_{rq}$, $\langle \xi^{\pm}_r | \xi^{\pm}_r \rangle = \langle \bar{\xi}^{\pm}_r | \bar{\xi}^{\pm}_r \rangle = 1$, and $\langle \xi^-_r | \xi^+_r \rangle = \langle \bar{\xi}^-_r | \bar{\xi}^+_r \rangle = \mathcal{R}_re^{i\theta_r}$ with $\mathcal{R}_r \le 1$. This is a general entangled state between the edge spin and the rest of the system with the edge spin constrained to satisfy $\langle S_1^x\rangle=0$. The structure of entanglement between them is encoded in the overlap properties of $|\xi^{\pm}_r \rangle$ and $|\bar{\xi}^{\pm}_r \rangle$, which will be highlighted in Sec~\ref{sec:EE}. 

These states can be utilized to encode the SZM operator with the leading order term dominated by $S_1^x$. 
\begin{equation}
  \hat{O}_{\text{szm}} = \sum_{r} \frac{e^{i\theta_r}}{2} |\Phi^+_r\rangle \langle\Phi^-_r| + \text{h.c.} \propto S^x_1 + \text{corrections},
  \label{eq:szm}
\end{equation}
The corrections to the leading order term are controlled by the extent of deviation from the complete overlap between $| \xi^+_r \rangle$ and $| \xi^-_r \rangle$ states. Despite the rapid thermalization in the bulk, on increasing the cluster of driven sites we find a proliferation of the number of Floquet eigenpairs satisfying the condition $\mathcal{R}_r \approx 1$, $\forall r$ (see Appendix~\ref{app:SZMmore}). Therefore, the form of the strong zero mode is indeed well-approximated by Eq.~\ref{eq:szm}. The existence of this normalizable operator explains the degeneracy present in the spectrum and relates it directly to the pairing of Floquet eigenstates.

\begin{figure}
    \centering
    \includegraphics[width=\linewidth]{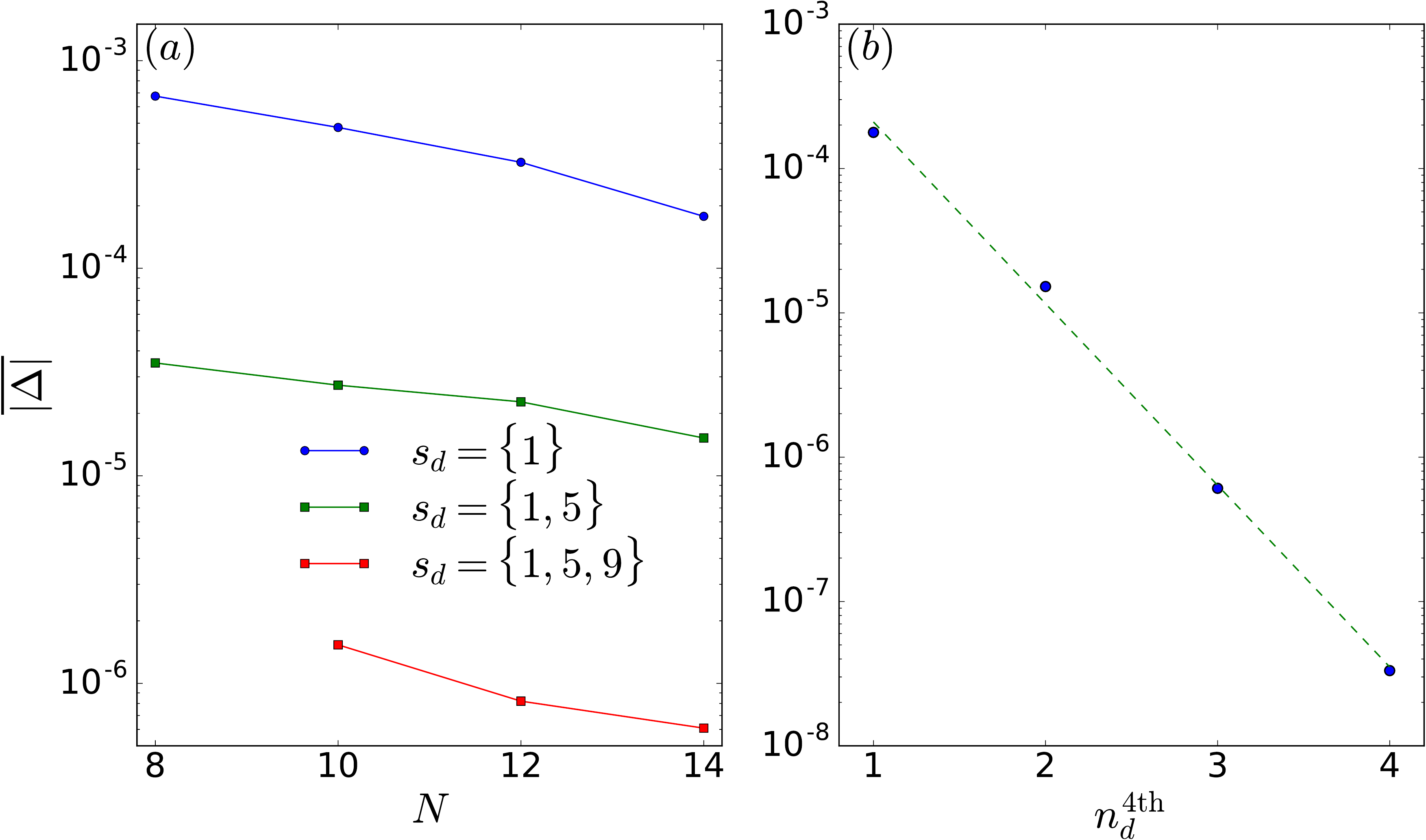}
    \caption{Scaling of averaged pairing gaps ($\overline{|\Delta|}$) (a) with system size ($N$) for a fixed number of driven sites and (b) with increasing the size of the driven cluster $n_d^\text{4th}$ for a fixed system size ($N=14$). 
    All other parameters are the same as in Fig.~\ref{fig:twosite}. The dashed green line in subfigure (b) is a best-fit curve to an exponential ansatz, $\overline{|\Delta|}\sim\exp(-2.9 n_d^\text{4th})$.}
    \label{fig:gapscaling}
\end{figure}

Finally, we turn our attention to the commutation properties of $\hat{O}_{\text{szm}}$ with the Floquet Hamiltonian. The commutator assumes the following form in terms of eigenstates and eigenvalues of $H_F$,
\begin{equation}
[H_F,\hat{O}_{\text{szm}}]=\sum_r (\epsilon^+_r - \epsilon^-_r) \frac{e^{i\theta_r}}{2} |\Phi^+_r\rangle \langle\Phi^-_r| - \text{h.c.},
\end{equation}
which implies that all spectral gaps must simultaneously vanish exponentially with $N$ for the SZM to be a quasi-conserved operator. In Fig.~\ref{fig:szm}(c) we present results for the average pairing gap 
\begin{equation}    
\overline{|\Delta|} = \sum_{r=1}^{2^{N-1}} |\epsilon^+_r - \epsilon^-_r| / 2^{N-1},
\end{equation} 
where indeed the gap decreases significantly with increasing size of the driven cluster ($n_d^\text{4th}$) in the proximity of the freezing frequency. Interestingly, when we fix $n_d^\text{4th}$ and start increasing the system size, the gap decreases weakly with $N$ [see Fig.~\ref{fig:gapscaling}(a)]. We associate this slow decay of $\Delta(N, n_d^{\text{4th}})$ with $N$ to the convergence properties of the corrections in Eq.~\eqref{eq:szm}, which appear to converge for smaller $N$, with $||[H_F,\hat{O}_{\text{szm}}]||$ decreasing in magnitude. Although, subsequently for larger $N$ the corrections dominate $\hat{O}_{\text{szm}}$ with the commutator reaching a finite threshold signalling the presence of an approximate SZM (as witnessed previously for the boundary driven case in Sec.\ref{sec:singlesite}). On the other hand, if the system size is fixed, the gap exhibits an \textit{exponential} fall with $n_d^\text{4th}$ as shown in Fig.~\ref{fig:gapscaling}(b) [see Appendix~\ref{app:semianalytical_gap} for more details]. 
This behavior of the pairing gap explains our results for the freezing dynamics of the boundary site in Fig.~\ref{fig:dynamics2}, where we found a similar dependence of the relaxation times on the system size and the driven cluster size. Therefore, we conclude that increasing both $N$ and $n_d^\text{4th}$ leads to an asymptotic exponential decrease of the average pairing gap and hence the commutator of $\hat{O}_{\text{szm}}$ with the Floquet Hamiltonian when $n_d^\text{4th}$ scales linearly with $N$.

In summary, we have demonstrated that the protocol involving driving every fourth site in the chain leads to the emergence of a SZM. We have proposed an approximate form of an operator that satisfies all the features of SZM which is directly related to the spectrum-wide quasi-degeneracy and slow relaxation rate of the boundary spin. We note that any bulk site can not be frozen in this manner which suggests that a strictly convergent SZM, localized in the bulk, can not be constructed.

\section{Athermal entanglement structure of Floquet eigenstates}
\label{sec:EE}

In this section, we discuss the non-equilibrium properties of the system through the lens of the entanglement entropy of Floquet eigenstates. The athermal and thermal behavior  represented in the dynamics and spectral properties of the boundary and bulk degrees of freedom respectively, have an imprint on the entanglement features of the eigenstates as well. In fully thermal eigenstates at infinite temperature, the reduced density matrix of any subsystem exhibits maximal entanglement. The disentangling nature of the periodic drive introduces several novel features in entanglement which we categorize between thermal and athermal depending on the partitioning of subsystems.

\begin{figure}
    \centering
    \includegraphics[width=\linewidth]{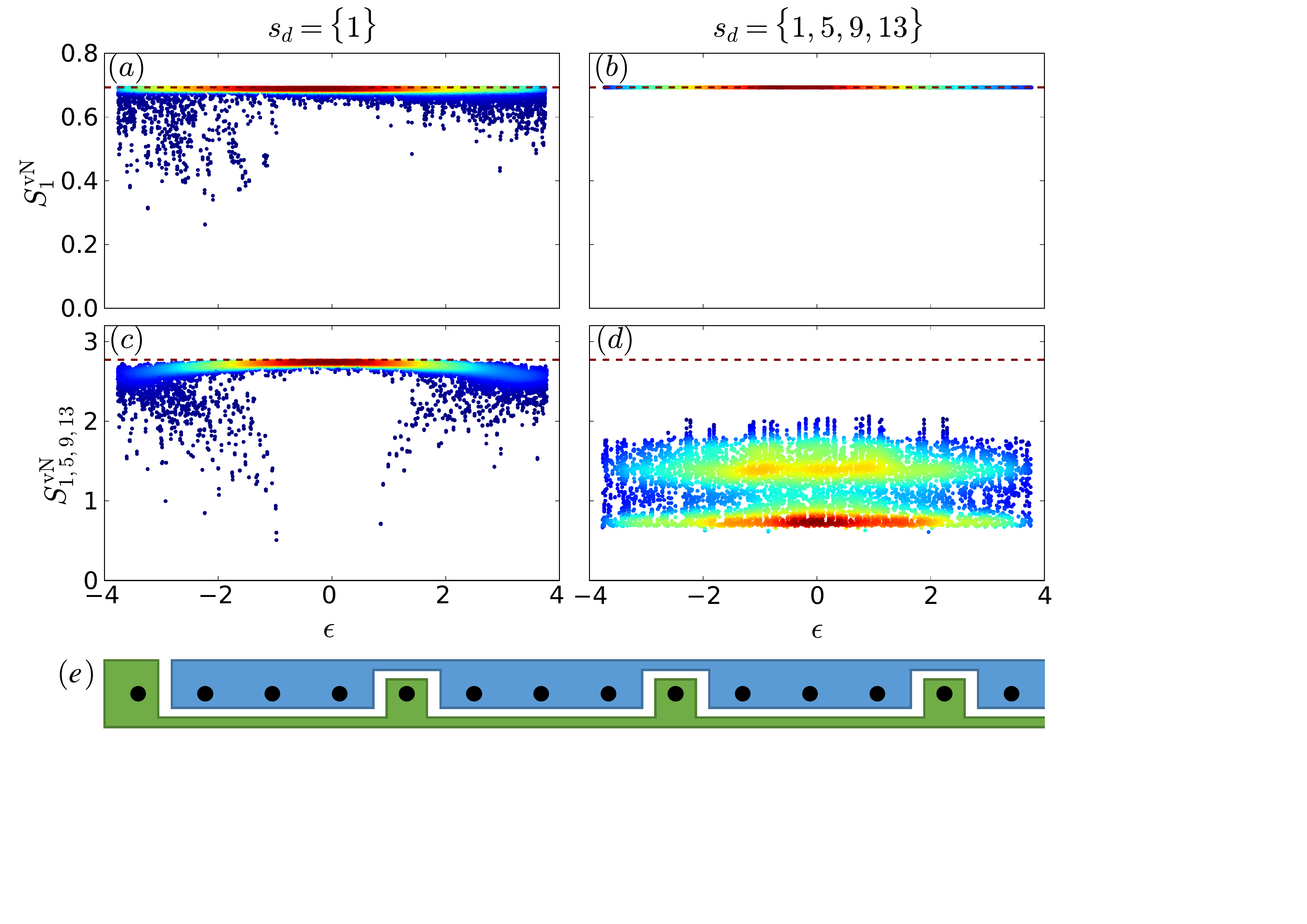}  
    \caption{Entanglement entropy of the boundary site with the rest of the system ($S^{\text{vN}}_1$) in the upper panels, and sites \{1,5,9,13\} with the rest of the system ($S^{\text{vN}}_{1,5,9,13}$) in the lower panels for $N=14$. The driven sites are mentioned at the top of each column. All other parameters are the same as in Fig.~\ref{fig:twosite}. Page values of entanglement entropies are shown by the red dashed line. The bipartition used to calculate the entanglement entropy in panels (c) and (d) is shown below in subfigure (e).}
    \label{fig:ent}
\end{figure}

The von Neumann entanglement entropy of the boundary site with the rest of the system in the Floquet eigenstates (Eq.\ref{eq:Floquet_eigenstates}) can be written as
\begin{equation}
  S^{\text{vN},\pm}_1 = -\frac{1+|\delta^\pm|}{2} \ln\frac{1+|\delta^\pm|}{2} - \frac{1-|\delta^\pm|}{2} \ln\frac{1-|\delta^\pm|}{2},
\end{equation}
where $\delta^\pm_r=\langle\xi^\pm_r|\bar{\xi}^\pm_r\rangle$.
In the protocol with $s_d=\{1\}$, $|\delta^{\pm}_r|$ is found to be large for many eigenstates, thus giving rise to many weakly entangled states, nonetheless in a large majority of states the edge spin remain maximally entangled with the bulk as shown in Fig.~\ref{fig:ent}(a). In the other limit, for $s_d=\{1,5,9,13\}$ the boundary spin is maximally entangled in \textit{all} the Floquet eigenstates [see Fig.~\ref{fig:ent}(b)] which also implies $|\delta^{\pm}_r|$ is vanishingly small. 
This is counterintuitive since, the enhanced freezing of the boundary spin is expected to lead to disentangling of the edge spin in the Floquet eigenstates, which instead appears fully ergodic for single-site observables. Therefore, the dynamics of the boundary spin, however slow it may be for finite systems, is only a prethermal phenomenon with the athermal behavior emerging in the thermodynamic limit. The boundary spin takes an exponentially long time in $N$ to dephase and finally reaches a featureless state for any finite system size.

However, the scaling of the Floquet eigenpair gap in Fig.~\ref{fig:gapscaling} suggests that it should vanish in the thermodynamic limit. Consequently, the decoherence time of the boundary spin diverges as well as the states $\ket{\Phi^+_r}$ and $\ket{\Phi^-_r}$ become degenerate which allows us to choose $(\ket{\Phi^+_r} \pm \ket{\Phi^-_r})/\sqrt{2}$ as the Floquet eigenstates, where the boundary spin is \textit{disentangled} from the rest of the system. Thus, the apparent paradox between the stroboscopic dynamics of the edge spin and the ergodic properties of the Floquet eigenstates is resolved in the thermodynamic limit.

Interestingly, we do find athermal entanglement signatures in the Floquet eigenstate when every fourth site is driven and the system is bipartitioned into driven and undriven sites [see Fig.~\ref{fig:ent}(e)]. For many eigenstates, the entanglement remains close to $\ln2$, as can be seen in Fig.~\ref{fig:ent}(d). Note that typical states in general should satisfy volume-law scaling for such bipartitions, an example being the protocol $s_d=\{1\}$ as shown in Fig.~\ref{fig:ent}(c). Therefore, our most efficient protocol ($s_d=\{1,3,5,9\}$) generates exponentially many Floquet eigenstates with \textit{subextensive} (athermal) entanglement, 
for a bipartition separating the driven and undriven sites. However, any random bipartition such as half-chain entanglement appears to produce a volume law scaling which is reminiscent of the behavior of rainbow states~\cite{Ramirez2015, Langlett2022}. 
Our entanglement results showcase a rich and complex structure based on the geometry of the driven sites, where we observe signatures of both prethermal and athermal behaviors.

\section{Conclusion and Discussion}
\label{sec:discussion}

The phenomena of SZM where a quasi-local conservation law localized on the boundary, is proven to exist in one-dimensional integrable models with a discrete symmetry. In non-integrable models, this operator usually develops a finite lifetime and ceases to be a conserved quantity in the thermodynamic limit. In this paper, we have shown the emergence of SZM in a nonintegrable spin chain through local Floquet engineering. We begin with a time-independent non-integrable Hamiltonian with a continuous $U(1)$ symmetry which is broken by the periodic drive. We consider two classes of protocol, one where only the boundary spin is driven while the multisite drive also includes a subset of spins in the bulk. Through a second-order F-M expansion of the Floquet Hamiltonian, we show that the edge spin can undergo a significant slowdown for the boundary drive at certain freezing frequencies, which arise due to the non-local corrections in the Floquet Hamiltonian. In contrast, such higher-order processes were previously reported to destroy the dynamic localization in interacting systems~\cite{verdeny}. At the freezing frequency, we argue that the system develops an approximate global discrete $\mathbb{Z}_2$ symmetry which is responsible for the boundary slowdown because of the formation of an approximate SZM. We find the boundary relaxation slows down with increasing system size but the relaxation time scale does not appear to grow exponentially. 

We propose a multi-site drive protocol that further slows down the boundary relaxation. A special sequence of driven sites leads to the slowdown which scales exponentially with the number of driven sites. The emergence of a SZM is shown through signatures in spectral properties and eigenstate overlaps. The significantly enhanced slowdown of the boundary spin coexists with the rapid thermalization of the bulk degrees of freedom which realizes a novel regime that exhibits thermal and athermal behavior. The intermediate athermal behavior also leaves an imprint in the entanglement structure of the Floquet eigenstates. Although the dynamics of the boundary spin relaxes on a time scale exponentially large in system size, the entanglement of the boundary spin is maximal in the Floquet eigenstates. Along with maximal single-site entanglement, random bipartitions also exhibit volume law scaling. Perhaps not surprisingly the bipartition involving the driven sites is significantly athermal. We leave the complete description of the entanglement structure as an interesting future problem. The eigenstates also show thermal and athermal properties reflecting the non-trivial consequences on the thermalization of the driven system. Our description provides a novel realization of an emergent SZM in a non-integrable system realized through local Floquet engineering. 

The stability of local athermal dynamics on a background of a thermalizing environment is of relevance for protecting quantum information. The emergence of quasi-local conservation laws with exponentially long relaxation times could potentially play a role near the many-body localization transition~\cite{Chandran2016, Kulshreshtha2018lbits}. The interplay between the local Floquet drive and emergent symmetries can provide a path towards realizing novel states with a rich entanglement structure far from the ground state. A general understanding of the entanglement structure of states which are generically volume-law entangled, yet contain elements of athermal character are potentially important for stabilizing non-equilibrium quantum orders away from the ground state. 
Our prescription for realizing local athermal dynamics utilizes the non-local multi-spin interactions generated by local Floquet engineering. This can be a valuable tool for engineering models with novel athermal character including QMBS~\cite{sanada2023QMBS}.  
The use of optical tweezers in Rydberg atom arrays opens the opportunity for implementing local drives where the physics of SZM can be investigated~\cite{Bluvstein, Martin2023LocalThermalization}.  Majorana zero modes in the ground state have been studied experimentally on solid state platforms \cite{nadj2014majorana, sarma2015majorana, lutchyn2018majorana}. Floquet driving provides another means of preparing such zero modes which can be robust to noise.

\begin{acknowledgments}
We thank P.\@ Sharma for collaboration on previous work. R.M.\@ and H.J.C.\@ acknowledge support from Grant No.\@ NSF DMR-2046570 and Florida State University (FSU) and the National High Magnetic Field Laboratory. The National High Magnetic Field Laboratory is supported by the National Science Foundation through Grant No.~DMR-1644779 and by the state of Florida. A.P.,  B. M., and M.S.\@ were funded by the European Research Council (ERC) under the European Union's Horizon 2020 research and innovation programme (Grant Agreement No.\@ 853368). We also thank the Planck cluster and the Research Computing Center (RCC) at FSU for computing resources. The authors acknowledge the use of the UCL Myriad High Performance Computing Facility (Myriad@UCL), and associated support services, in the completion of this work.
\end{acknowledgments}

\appendix
\section{Calculation of the Floquet Hamiltonian for single (edge) site driving}
\label{app:A}

Here we calculate the Floquet Hamiltonian $H_F$ for a boundary-driven protocol, where a single site at the edge is driven. We first define the Hamiltonian and the drive protocol 
\begin{equation}
    H(t)=H_0+f(t)V,
\end{equation}
where $H_0 = \sum_{i=1}^{N-1} (-1)^{i} {\bf S}_i \cdot {\bf S}_{i+1} - h \sum_{i=1}^N S^z_i$, $f(t)=\gamma \sgn(\sin(\omega t))$ and $V = S^x_1$, and we are using open boundary conditions.

We first apply a rotating-wave transformation to the time-dependent Hamiltonian,
\begin{equation}
    H_{\text{rot}}(t)=W^{\dagger}H(t)W-i W^{\dagger}\partial_t W,
\end{equation}
where 
\begin{equation}
    W(t)=e^{-i \int_0^t dt' f(t') V}=e^{-i \theta(t) S^x_1},
\end{equation}
with $\theta(t) = \gamma t\sum_n[\Theta(n\tau+\tau/2-t)\,\Theta(t-n\tau)]+\gamma(\tau-t)\sum_n[\Theta((n+1)\tau-t)\,\Theta(t-(n\tau+\tau/2))]$, where $\Theta$ is a Heviside step function. Firstly, this eliminates the explicit presence of the driving term in the time-dependent Hamiltonian, and secondly, $H_F$ obtained in this way yields a resummed expression in $\omega$ (which increases the radius of convergence of the series) and can be considered as a perturbative expansion in $1/\gamma$ only.
This allows us to consider all possible $\omega$ but restrict ourselves to a high drive amplitude regime.
We obtain
\begin{align}
    H_{\text{rot}}(t)&=W^{\dagger}H_0W\nonumber\\
    &=-S^x_1S^x_2-\cos\theta(S^y_1S^y_2+S^z_1S^z_2)\nonumber\\
    &\quad +\sin\theta(S^z_1S^y_2-S^y_1S^z_2)-h(\cos\theta S^z_1+\sin\theta S^y_1) \nonumber \\
    &\quad + \sum_{i=2}^{N-1}(-1)^{i}{\bf S}_i.{\bf S}_{i+1}-h\sum_{i=2}^{N}S^z_i,
\end{align}
where we have used $W^{\dagger}S^z_1W=\cos(\theta) S^z_1+\sin(\theta) S^y_1$ and $W^{\dagger}S^y_1W=-\sin(\theta) S^z_1 + \cos(\theta) S^y_1$. For brevity, we replace $\theta(t)$ by $\theta$.

The stroboscopic Floquet Hamiltonian in the F-M expansion is given by 
\begin{equation}
    H_F=\sum_{l=0}^{\infty}H_F^{(l)}.
\end{equation}
We calculate this expression analytically up to the 2nd term ($l=2$) in the subsections below.

\subsubsection{$l=0$ term in F-M}

The $l=0$ Floquet Hamiltonian is just the time-averaged $H_{\text{rot}}(t)$ over one time period,
\begin{align}
    H_F^{(0)}&=\frac{1}{\tau}\int_0^\tau H_{\text{rot}}(t)dt\nonumber\\
  &=-S^x_1S^x_2-\frac{h}{\lambda}[\sin(\lambda) S^z_1+(1-\cos\lambda)S^y_1]\nonumber\\
  &\quad -\frac{\sin\lambda}{\lambda}(S^y_1S^y_2+S^z_1S^z_2)+\frac{1{-}\cos\lambda}{\lambda}(S^z_1S^y_2-S^y_1S^z_2)\nonumber\\
  &\quad+\sum_{i=2}^{N-1}(-1)^{i}\,{\bf S}_i\cdot{\bf S}_{i+1}-h\sum_{i=2}^{N}S^z_i.
\end{align}
We have used the following integrals: $I_1[\cos\theta] = 2\sin(\lambda)/\gamma$, $I_1[\sin\theta] = 2(1-\cos\lambda)/\gamma$, where $I_1[x]\equiv\int_0^\tau x\, dt$ and $\lambda=\gamma \tau/2$. The form of this Hamiltonian suggests that if we choose the drive parameters accordingly ($\gamma \tau = 4 k \pi$) then $S^x_1$ is stroboscopically conserved, as $[S^x_1, H_F^{(0)}] = 0$  in this case.

\subsubsection{$l=1$ term in F-M}

The first term in F-M is given by
\begin{equation}
H_F^{(1)}=\frac{1}{2!i\tau}\int_0^\tau dt_1\int_0^{t_1}dt_2[H_{\text{rot}}(t_1),H_{\text{rot}}(t_2)].
\end{equation}
We obtain,
\begin{align}
 &[H_{\text{rot}}(t_1),H_{\text{rot}}(t_2)] =\nonumber\\
 &\qquad= i \text{s}_{12} \left\{S_1^{x} (2 h^2+4 h S_2^{z}+1) \text{c}_{12} - 2 S_1^{y} \left[2 h S_2^{x} \text{s}_{12} \right. \right. \nonumber\\
 &\qquad - S_2^{x} S_3^{y} \text{c}_{12} - S_2^{x} S_3^{z} \text{s}_{12} + S_3^{x} S_2^{y} \text{c}_{12} + S_3^{x} S_2^{z} \text{s}_{12} ]  \nonumber\\
 &\qquad - 2 S_1^{z} [2 h S_2^{x}\text{c}_{12} + S_2^{x} S_3^{y} \text{s}_{12} - S_2^{x} S_3^{z} \text{c}_{12} - S_3^{x} S_2^{y} \text{s}_{12} \nonumber\\
 &\qquad \left. \left. + S_3^{x} S_2^{z} \text{c}_{12} \right] -S_2^{x} \text{c}_{12} \right\},
\end{align}
where we denote $\theta(t_1)=\theta_1$, $\theta(t_2)=\theta_2$, and use a short-hand notation $\text{s}_{12} \equiv \sin[(\theta_1 - \theta_2)/2]$ and $\text{c}_{12} \equiv \cos[(\theta_1 - \theta_2)/2]$. By performing the integrals, we find 
\begin{equation}
    H_F^{(1)} = 0.
\end{equation}
Here we have used: $I_2[\sin(\theta_1 - \theta_2)] = 0$, $I_2[\cos\theta_1] = I_2[\cos\theta_2] = \tau\sin(\lambda)/\gamma$, $I_2[\sin\theta_1] 
= I_2[\sin\theta_2] = \tau(1-\cos\lambda)/\gamma$ where $I_2[x] = \int_0^\tau dt_1 \int_0^{t_1} dt_2 x$. Thus we find that the 1st term in F-M is zero.

\subsubsection{$l=2$ term in F-M}

The 2nd term in F-M is given by
\begin{align}
    H_F^{(2)}&=\frac{1}{3!\tau i^2}\int_0^\tau dt_1\int_0^{t_1}dt_2\int_0^{t_2}dt_3\nonumber\\
    &\quad\times([H_{\text{rot}}(t_1),[H_{\text{rot}}(t_2),H_{\text{rot}}(t_3)]]+(1\leftrightarrow3)).
    \label{2ndorder}
\end{align}
Calculation of these higher-order nested commutators is straightforward but unwieldy. In this section, we will consider only the renormalization of $H_F^{(0)}$ i.e.\@ only the one- and the two-site terms which involve spin operators at the driven (boundary) site, more specifically only $S^y_1$ and $S^z_1$ which does not commute with $S^x_1$. We designate this part $H^{(2)}_{F,\ \text{loc}}$, and show that even the $\mathcal{O}(1/\gamma)$ terms present in it is sufficient to demonstrate the absence of dynamic freezing of the edge spin. Later we will also consider the other terms which are mostly long-range in nature ($H^{(2)}_{F,\ \text{nonloc}}$) and responsible for the enhanced freezing of the boundary spin. So, $H_F^{(2)} = H^{(2)}_{F,\ \text{loc}} + H^{(2)}_{F,\ \text{nonloc}}$.

Now, we have
\begin{align}
   &\{[H_{\text{rot}}(t_1),[H_{\text{rot}}(t_2),H_{\text{rot}}(t_3)]]+(1\leftrightarrow3)\}_{\text{loc}}=\nonumber\\
   &\quad= A_1(\boldsymbol{\theta})S^z_1 + A_2(\boldsymbol{\theta})S^y_1 + A_3(\boldsymbol{\theta})S^z_1S^z_2 + A_4(\boldsymbol{\theta})S^y_1S^y_2\nonumber\\
   &\qquad + A_5(\boldsymbol{\theta})S^z_1S^y_2 + A_6(\boldsymbol{\theta})S^y_1S^z_2 + \cdots,
   \label{eq:Hrot1}
\end{align}
where $\boldsymbol{\theta}\equiv(\theta_1,\theta_2,\theta_3)$. As one can see, these terms are exactly the same as appearing in the $l=0$ term, thus only renormalizing their strength.

The coefficients are:
{\allowdisplaybreaks
\begin{align}
    A_1(\boldsymbol{\theta}) &= -\frac{h}{2}[\text{c}_1+\text{c}_3-2\text{c}_2\nonumber\\*
    &\quad+2(1+h^2)(\text{s}_1\text{s}_2\text{c}_3 + \text{c}_1\text{s}_2\text{s}_3 - 2\text{s}_1\text{c}_2\text{s}_3)],
    \label{eq:A1}\\
    A_2(\boldsymbol{\theta}) &= -\frac{h}{2}[\text{s}_1+\text{s}_3-2\text{s}_2\nonumber\\*
    &\quad+2(1+h^2)(\text{s}_1\text{c}_2\text{c}_3 +     \text{c}_1\text{c}_2\text{s}_3 - 2\text{c}_1\text{s}_2\text{c}_3)],
    \label{eq:B1}\\
    A_3(\boldsymbol{\theta}) &= -\frac{1}{2}[\text{c}_1+\text{c}_3-2\text{c}_2\nonumber\\*
    &\quad+2(1+3 h^2) (\text{s}_1\text{s}_2\text{c}_3 + \text{c}_1\text{s}_2\text{s}_3 - 2\text{s}_1\text{c}_2\text{s}_3)],
    \label{eq:C1}\\
    A_4(\boldsymbol{\theta}) &= -\frac{1}{2}[(1+4h^2)(\text{c}_1+\text{c}_3-2\text{c}_2)\nonumber\\*
    &\quad+2(1+h^2)  (\text{s}_1\text{s}_2\text{c}_3 + \text{c}_1\text{s}_2\text{s}_3 - 2\text{s}_1\text{c}_2\text{s}_3)],
    \label{eq:D}\\
    A_5(\boldsymbol{\theta}) &= \frac{1}{2}[(1+4h^2)(\text{s}_1+\text{s}_3-2\text{s}_2)\nonumber\\*
    &\quad+2(1+h^2)(\text{s}_1\text{c}_2\text{c}_3 + \text{c}_1\text{c}_2\text{s}_3-2\text{c}_1\text{s}_2\text{c}_3)],
    \label{eq:E1}\\
    A_6(\boldsymbol{\theta}) &= -\frac{1}{2}[\text{s}_1+\text{s}_3-2\text{s}_2\nonumber\\*
    &\quad+2(1+3h^2)(\text{s}_1\text{c}_2\text{c}_3 +   \text{c}_1\text{c}_2\text{s}_3 - 2\text{c}_1\text{s}_2\text{c}_3)],
  \label{eq:F1}
\end{align}
}where we use a short-hand notation $\text{s}_i \equiv \sin\theta_i, \text{c}_i \equiv \cos\theta_i$. This yields the following integrals ($I_3[x] \equiv \int_0^\tau dt_1 \int_0^{t_1} dt_2 \int_0^{t_2} dt_3 x$):
{\allowdisplaybreaks
\begin{align}
  I_3[A_1(\boldsymbol{\theta})] &= \tfrac{12h \gamma \tau (h^2+1) \cos \lambda + h (\tau^2 \gamma^2-8 h^2+16) \sin \lambda}{4\gamma^3}\nonumber\\*
  &\quad-\tfrac{8 h (h^2+1) \sin2\lambda + 12 h \gamma \tau}{4\gamma^3},
  \label{eq:IntA1}\\
  I_3[A_2(\boldsymbol{\theta})] &= \tfrac{8h(2 h^2-1) \cos \lambda+8h(h^2+1) \cos2\lambda-24h^3}{4\gamma^3}\nonumber\\*
  &\quad+\tfrac{12h\gamma \tau (h^2+1) \sin \lambda-h\gamma^2\tau^2 (\cos \lambda+2)}{4\gamma^3},
  \label{eq:IntB1}\\
  I_3[A_3(\boldsymbol{\theta})] &= \tfrac{(-24 h^2+\tau^2 \gamma^2+16) \sin \lambda-8 (3 h^2+1) \sin2\lambda}{4 \gamma^3}\nonumber\\*
  &\quad+\tfrac{12 (3 h^2+1) \gamma \tau \cos \lambda-12 \gamma \tau}{4 \gamma^3},
  \label{eq:IntC1}\\
  I_3[A_4(\boldsymbol{\theta})] &= \tfrac{[(4 h^2+1) \tau^2 \gamma^2+88 h^2+16] \sin \lambda +12 (h^2+1) \gamma \tau \cos \lambda}{4 \gamma^3}\nonumber\\*
  &\quad-\tfrac{12 (4 h^2+1) \gamma \tau+8 (h^2+1) \sin2\lambda}{4 \gamma^3},
  \label{eq:D1}\\
  I_3[A_5(\boldsymbol{\theta})] &= \tfrac{8(10 h^2+1) \cos \lambda-8 (h^2+1) \cos2\lambda -72 h^2}{4 \gamma^3}\nonumber\\*
  &\quad+ \tfrac{(4 h^2+1) \gamma^2\tau^2 (\cos \lambda+2) - 12\gamma \tau (h^2+1) \sin \lambda}{4 \gamma^3},
  \label{eq:IntE1}\\
  I_3[A_6(\boldsymbol{\theta})] &= \tfrac{12\gamma \tau (3 h^2+1)\sin\lambda-\gamma^2\tau^2(\cos\lambda+2)}{4 \gamma^3}\nonumber\\*
  &\quad-\tfrac{16 \sin ^2(\frac{\lambda}{2})[2(3 h^2+1) \cos\lambda + 12 h^2+1]}{4 \gamma^3},
\label{eq:IntF1}
\end{align}
}where we have used the values of the integrals,
{\allowdisplaybreaks
\begin{align}
  I_3[\text{c}_1]&=I_3[\text{c}_3]=\frac{1}{4\gamma^3}[4\gamma \tau+(\sin\lambda)(\gamma^2\tau^2-8)],\nonumber\\
  I_3[\text{c}_2]&=\frac{1}{2\gamma^3}[-4\gamma \tau+(\sin\lambda)(\gamma^2\tau^2+8)],\nonumber\\
  I_3[\text{s}_1]&=I_3[\text{s}_3]=\frac{1}{4\gamma^3}[2\gamma^2\tau^2-(\cos\lambda)(\gamma^2\tau^2-8)-8],\nonumber\\
  I_3[\text{s}_2]&=\frac{1}{2\gamma^3}\left[8 \sin^2\frac{\lambda}{2}-(\cos\lambda)(\gamma^2\tau^2+4)+4\right],
  \label{integrals1}
\end{align}
}and
{\allowdisplaybreaks
\begin{align}
&I_3[\text{s}_1\text{c}_2\text{c}_3]=\tfrac{10-6\cos\lambda-6\cos2\lambda+2\cos3\lambda-6\lambda\sin\lambda}{6\gamma^3},\nonumber\\
&I_3[\text{c}_1\text{s}_2\text{c}_3]=\tfrac{-4+3\cos\lambda+\cos3\lambda+6\lambda\sin\lambda}{3\gamma^3},\nonumber\\
&I_3[\text{c}_1\text{c}_2\text{s}_3]=\tfrac{10-6\cos\lambda-6\cos2\lambda+2\cos3\lambda-6\lambda\sin\lambda}{6\gamma^3},\nonumber\\
&I_3[\text{s}_1\text{c}_2\text{s}_3]=\tfrac{3\sin\lambda-6\sin2\lambda+\sin3\lambda+6\lambda\cos\lambda}{3\gamma^3},\nonumber\\
&I_3[\text{s}_1\text{s}_2\text{c}_3]=\tfrac{12\sin\lambda-6\sin2\lambda+2\sin3\lambda-6\lambda\cos\lambda}{6\gamma^3},\nonumber\\
&I_3[\text{c}_1\text{s}_2\text{s}_3]=\tfrac{12\sin\lambda-6\sin2\lambda+2\sin3\lambda-6\lambda\cos\lambda}{6\gamma^3}.
\label{integrals2}
\end{align}
}Values of these integrals substituted in Eq.~\eqref{2ndorder} directly give the $H^{(2)}_{F,\text{loc}}$. The interesting fact is that even $H^{(2)}_{F,\text{loc}}$ contains terms which are $\mathcal{O}(1/\gamma)$. This modifies the freezing conditions which were believed to be true at least in $\mathcal{O}(1/\gamma)$.

\subsubsection{Full $H_F$ (local)}

As mentioned before, the resummed $H_F$ is a series expansion in $1/\gamma$ and for many purposes, we can neglect the higher order (which are at least $1/\gamma^2$ or smaller) terms.
Therefore, summing up all the one and two site terms up to 2nd order in F-M expansion which are $\mathcal{O}(1/\gamma)$, we get the local Floquet effective Hamiltonian [Eq.~\eqref{eq:HF2_1overgamma} in the main text]
\begin{widetext}
\begin{align}
H_{F,\text{loc}}[\mathcal{O}(\frac{1}{\gamma})]&=H_F^{(0)}+H_F^{(1)}+H^{(2)}_{F,\text{loc}}[\mathcal{O}(1/\gamma)]+\cdots\nonumber\\
&=-S^x_1S^x_2-\frac{2\sin\lambda}{\gamma \tau}\left[\left(1+\frac{(1+4h^2)\tau^2}{48}\right) S^y_1S^y_2 + \left(1+\frac{\tau^2}{48}\right) S^z_1S^z_2\right]\nonumber\\
&\quad+\frac{2}{\gamma \tau}\left[\left(1-\cos\lambda -\frac{(1+4h^2)\tau^2(2+\cos\lambda)}{48}\right) S^z_1S^y_2
-\left(1-\cos\lambda-\frac{\tau^2(2+\cos\lambda)}{48}\right) S^y_1S^z_2 \right]\nonumber\\
&\quad-\frac{2h}{\gamma \tau}\left[(\sin\lambda)\left(1+\frac{\tau^2}{48}\right) S^z_1+\left(1-\cos\lambda-\frac{\tau^2(2+\cos\lambda)}{48}\right) S^y_1 \right] +
\sum_{i=2}^{N-1}(-1)^{i}{\bf S}_i\cdot{\bf S}_{i+1}-h\sum_{i=2}^{N}S^z_i.
\label{eq:HFlocal}
\end{align}

For concreteness, we also write the full local $H_F$ (considering the full contribution from the $l=2$ term in F-M expansion)
\begin{align}
  H_{F,\text{loc}}[\mathcal{O}(\frac{1}{\gamma^3})] &= H_F^{(0)}+H_F^{(1)}+H^{(2)}_{F,\ \text{loc}}[\mathcal{O}(\frac{1}{\gamma^3})]\nonumber\\
  &= -S^x_1S^x_2 - \left[\left(\frac{\sin\lambda}{\lambda}-a_4(\boldsymbol{p})\right) S^y_1S^y_2 + \left(\frac{\sin\lambda}{\lambda} - a_3(\boldsymbol{p})\right) S^z_1S^z_2 \right] + \left[\left(\frac{1-\cos\lambda}{\lambda}+a_5(\boldsymbol{p})\right) S^z_1S^y_2 \right.
  \nonumber\\
  &\quad\left.-\left(\frac{1-\cos\lambda}{\lambda}-a_6(\boldsymbol{p})\right) S^y_1S^z_2 \right] - \left[\left(\frac{h\sin\lambda}{\lambda}-a_1(\boldsymbol{p})\right) S^z_1 + \left(\frac{h(1-\cos\lambda)}{\lambda}-a_2(\boldsymbol{p})\right) S^y_1 \right]\nonumber\\
&\quad+\sum_{i=2}^{N-1}(-1)^{i}{\bf S}_i\cdot{\bf S}_{i+1}-h\sum_{i=2}^{N}S^z_i,
\label{eq:fullHFlocal}
\end{align}
\end{widetext}
where $a_i(\boldsymbol{p}) = - I_3[A_i(\boldsymbol{\theta})] / (6\tau)$.

\begin{table*}[tb]
{\renewcommand{\arraystretch}{2.0}
\centering
\begin{tabular}{ p{2.3cm} m{4.5cm} m{10.0cm} }
\hline \hline
 & $H_F^{(0)} + H_F^{(1)} + H_{F,\text{loc}}^{(2)}[\mathcal{O}(\frac{1}{\gamma})]$ & $H_F^{(0)} + H_F^{(1)} + H^{(2)}_{F,\ \text{loc}}[\mathcal{O}(\frac{1}{\gamma^3})]$ \\
\hline
Re[$H_F(2,1)$] & $-\frac{h\sin\lambda}{\gamma \tau}(1+\frac{\tau^2}{48})$ & $\frac{8(h+h^3)\sin2\lambda+12h\gamma \tau-12\gamma \tau(h+h^3)\cos\lambda+h\sin\lambda[8(h^2-2)-(\tau^2+48)\gamma^2]}{48\tau\gamma^3}$\\%\cline{2-3}

Im[$H_F(2,1)$]  & $-\frac{h}{\gamma \tau}(1-\cos\lambda)+
\frac{h\tau(2+\cos\lambda)}{48\gamma}$ & $\frac{24h^3+2h\gamma^2(\tau^2-24)-8(h+h^3)\cos2\lambda-12\gamma \tau(h+h^3)\sin\lambda+h\cos\lambda[8-16h^2+
(\tau^2+48)\gamma^2]}{48\tau\gamma^3}$\\%\cline{2-3}
\hline \hline
\end{tabular}
\caption{Analytical expressions of the matrix element $H_F(2,1)$ from F-M expansion.}
\label{tab:HF12}
}
\end{table*}

To demonstrate the accuracy of the resummed F-M expansion, we compute the matrix element $H_F(2,1)$ using Eq.~\ref{eq:HFlocal}, Eq.~\ref{eq:fullHFlocal} and summarize the consolidated expression in Table~\ref{tab:HF12}. We compare these expressions with exact numerics in Fig.~\ref{fig:matching} in the main text.

\subsubsection{Additional (long range) terms in $H_F^{(2)}$ : $H^{(2)}_{F,\ \text{nonloc}}$}

Here we consider additional terms in $H_F^{(2)}$ which are long-range (i.e.\@ beyond nearest neighbor) and multispin in nature which enhances the freezing of the boundary site. These terms are expected to be perturbatively suppressed for global driving, keeping the quasi-local nature of $H_F$ intact. Here, we will show that for local driving these terms can be similar in strength to the zeroth order terms. 

\begin{align}
  &\{[H_{\text{rot}}(t_1),[H_{\text{rot}}(t_2),H_{\text{rot}}(t_3)]]+(1\leftrightarrow3)\}_{\text{nonloc}}=\nonumber\\
  &\quad = \beta(\boldsymbol{\theta})(3S^z_1S^z_3 + 4S^z_1S^y_2S^y_3S^z_4 - 4S^z_1S^y_2S^z_3S^y_4\nonumber\\
  &\qquad - 4S^z_1S^x_2S^z_3S^x_4 + 4S^z_1S^x_2S^x_3S^z_4 + 3S^y_1S^y_3\nonumber\\
  &\qquad + 12hS^y_1S^y_2S^z_3S^y_4 - 4S^y_1S^z_2S^z_3S^y_4 - 12hS^y_1S^z_2S^y_3\nonumber\\
  &\qquad - 4S^y_1S^z_2S^y_3S^z_4 - 4S^y_1S^x_2S^y_3S^x_4 + 4S^y_1S^x_2S^x_3S^y_4)\nonumber\\
  &\qquad + \eta(\boldsymbol{\theta})(3S^z_1S^y_3 - 12hS^z_1S^y_2S^z_3 - 4S^z_1S^z_2S^z_3S^y_4\nonumber\\
  &\qquad + 12hS^z_1S^z_2S^y_3 + 4S^z_1S^z_2S^y_3S^z_4 + 4S^z_1S^x_2S^y_3S^x_4\nonumber\\
  &\qquad - 4S^z_1S^x_2S^x_3S^y_4 + 3S^y_1S^z_3 + 4S^y_1S^y_2S^y_3S^z_4\nonumber\\
  &\qquad + 4S^y_1S^y_2S^z_3S^y_4 + 4S^y_1S^x_2S^z_3S^x_4 - 4S^y_1S^x_2S^x_3S^z_4) + \cdots,
\label{longrange}
\end{align}
where the ``$\cdots$'' represents some terms which commute with $S^x_1$. The coefficients and their integrals are given by
{\allowdisplaybreaks
\begin{align}
  \beta(\boldsymbol{\theta})&=\frac{1}{4}[\cos\theta_1+\cos\theta_3-2\cos\theta_2],\nonumber\\
  I_3[\beta(\boldsymbol{\theta})]&=\frac{12 \tau \gamma-(\tau^2 \gamma^2+24) \sin\lambda}{8 \gamma^3},\nonumber\\
  \eta(\boldsymbol{\theta})&=\frac{1}{4}[\sin\theta_1+\sin\theta_3-2\sin\theta_2],\nonumber\\
  I_3[\eta(\boldsymbol{\theta})]&=\frac{\tau^2 \gamma^2 (\cos\lambda+2)-48 \sin^2\frac{\lambda}{2}}{8 \gamma^3}.
\label{integrals3}
\end{align}
}Substituting Eq.~\eqref{longrange} and the values of the integrals in Eq.~\eqref{integrals3} to Eq.~\eqref{2ndorder}, we directly get the non-local part of $H_F^{(2)}$. The crucial thing to note here is that, even at the special frequencies ($\gamma \tau = 4n\pi$), some of the integrals in Eq.~\eqref{integrals3} (which determines the strength of the long-range terms) are not only nonzero, but similar in magnitude to the zeroth order terms i.e.\@ $\mathcal{O}(1/\gamma)$. Counter-intuitively, as shown in the main text, these terms take part in enhancing the freezing of the boundary spin at $\omega^k_m$.

\section{Calculation of $H_F$ for two-site driving}
\label{app:B}

Here we give a detailed derivation of the Floquet Hamiltonian for two-site driving i.e.\@ when a second site (say, site $j$) is driven in addition to the boundary site (site 1) following the same drive protocol.
The time-dependent part of the Hamiltonian is $\gamma \sgn(\sin(\omega t))(S^x_1+S^x_j)$ and we follow the same prescription of calculating $H_F$ as charted out in the previous section for single site driving. The rotating wave transformation is generated by the operator: $W(t)=e^{-i\theta(t)(S^x_1+S^x_j)}$ which gives for $j>2$
\begin{align}
    &H^{j>2}_{\text{rot}}(t) = W^{\dagger}H_0W\nonumber\\
    &=-S^x_1S^x_2-\cos\theta(S^y_1S^y_2+S^z_1S^z_2)+\sin\theta(S^z_1S^y_2-S^y_1S^z_2)\nonumber\\
    &\quad+(-1)^{j-1}[S^x_{j-1}S^x_j+\cos\theta(S^y_{j-1}S^y_j+S^z_{j-1}S^z_j)\nonumber\\
    &\quad-\sin\theta(S^z_{j-1}S^y_j-S^y_{j-1}S^z_j)]\nonumber\\
    &\quad+(-1)^j[S^x_{j}S^x_{j+1} + \cos\theta(S^y_jS^y_{j+1}+S^z_jS^z_{j+1})\nonumber\\
    &\quad-\sin\theta(S^z_jS^y_{j+1}-S^y_jS^z_{j+1})]\nonumber\\
    &\quad-h[\cos\theta (S^z_1+S^z_j)+\sin\theta(S^y_1+ S^y_j)]\nonumber\\
    &\quad+\sum_{i=2}^{j-2}(-1)^{i}{\bf S}_i\cdot{\bf S}_{i+1}+\sum_{i=j+1}^{N-1}(-1)^{i}{\bf S}_i\cdot{\bf S}_{i+1}\nonumber\\
    &\quad-h\left(\sum_{i=2}^{j-1}S^z_i+\sum_{i=j+1}^NS^z_i\right),
\end{align}
and for $j=2$
\begin{align}
    H^{j=2}_{\text{rot}}(t) &= W^{\dagger}H_0W\nonumber\\
    &=-{\bf S}_1\cdot{\bf S}_2+S^x_2S^x_3\nonumber\\
    &\quad+\cos\theta(S^y_2S^y_3+S^z_2S^z_3)-\sin\theta(S^z_2S^y_3-S^y_2S^z_3)\nonumber\\
    &\quad-h[\cos\theta (S^z_1+S^z_2)+\sin\theta(S^y_1+ S^y_2)]\nonumber\\
    &\quad+\sum_{i=3}^{N-1}(-1)^{i}{\bf S}_i\cdot{\bf S}_{i+1}-h\sum_{i=3}^{N}S^z_i.
\end{align}

Note that, $SU(2)$ symmetry is locally restored for the $j=2$ case as the interaction term ${\bf S}_1\cdot{\bf S}_2$ is fully rotationally invariant. This reduces the number of additional terms generated by the drive and leads to a significant suppression of the matrix element $H_F(4,1)$ (where $\ket{4}=\sigma^z_1\sigma^z_2\ket{1}$) which in turn reduces the rate of relaxation of the spins at sites 1 and 2. We focus on calculating $H_F$ for $s_d=\{1,5\}$ in the rest of this Appendix. 

Proceeding in the same manner as before, we get the $l=0$ term in F-M,
\begin{widetext}
\begin{align}
  H_F^{(0)}=\frac{1}{\tau}\int_0^\tau H_{\text{rot}}^{j=5}(t)dt
  &=-S^x_1S^x_2-\frac{\sin\lambda}{\lambda}(S^y_1S^y_2+S^z_1S^z_2)+\frac{1-\cos\lambda}{\lambda}(S^z_1S^y_2-S^y_1S^z_2) - \frac{h}{\lambda}[\sin\lambda S^z_1+(1-\cos\lambda)S^y_1]\nonumber\\
  &\quad+S^x_4S^x_5+\frac{\sin\lambda}{\lambda}(S^y_4S^y_5+S^z_4S^z_5)-\frac{1-\cos\lambda}{\lambda}(S^z_4S^y_5-S^y_4S^z_5) - \frac{h}{\lambda}[\sin\lambda S^z_5+(1-\cos\lambda)S^y_5]\nonumber\\
  &\quad-S^x_5S^x_6-\frac{\sin\lambda}{\lambda}(S^y_5S^y_6+S^z_5S^z_6)+\frac{1-\cos\lambda}{\lambda}(S^z_5S^y_6-S^y_5S^z_6)\nonumber\\
  &\quad+{\bf S}_2\cdot{\bf S}_3-{\bf S}_3\cdot{\bf S}_4+\sum_{i=6}^{N-1}(-1)^{i}{\bf S}_i\cdot{\bf S}_{i+1}-h\left(S^z_2+S^z_3+S^z_4+\sum_{i=6}^NS^z_i\right).
\end{align}
\end{widetext}
We again find, $H_F^{(1)} = 0$.

We notice that all local terms (one and two-body operators) in $H_F^{(2)}$ involving the spin operators at site 1 ($S^x_1,S^y_1,S^z_1$) are the same as for the case of boundary driving ($s_d=\{1\}$) given in Eq.~\eqref{eq:Hrot1}--\eqref{eq:HFlocal} which means they are not affected by the drive at site 5. 
We do not attempt to calculate the full $H_F^{(2)}$ in this case as we find that it is not sufficient to have good matching with exact numerical data even for very simple events like a single spin flip at site 5 from the state $\ket{\psi_0}$ determined by the matrix element $H_F(17,1)$ ($\ket{1}=\ket{\psi_0}$ and $\ket{17}=\sigma^z_5\ket{\psi_0}$).
To this end, we find the following terms in $H_F^{(2)}$ which contribute to $H_F(17,1)$,
\begin{align}
  &\{[H_{\text{rot}}^{j=5}(t_1),[H_{\text{rot}}^{j=5}(t_2),H_{\text{rot}}^{j=5}(t_3)]]+(1\leftrightarrow3)\}_{\text{local}}=\nonumber\\
  &\quad=\cdots + A(\boldsymbol{\theta})S^z_5 + B(\boldsymbol{\theta})S^x_4S^z_5S^x_6\nonumber\\
  &\qquad+ C(\boldsymbol{\theta})S^y_5 + D(\boldsymbol{\theta})S^x_4S^y_5S^x_6 + \cdots,
  \label{eq:Hrot15}
\end{align}
where the first ``$\cdots$'' in R.H.S.\@ stands for the same terms as in Eq.~\eqref{eq:Hrot1} and the last ``$\cdots$'' represent additional terms generated due to the driving of site 5.
Note that unlike the driven boundary site, three body interaction terms are also contributing to flip the spin at the driven site 5 in addition to the local field terms ($S^z_5$, $S^y_5$). The coefficients of the terms in Eq.~\eqref{eq:Hrot15} are given by
{\allowdisplaybreaks
\begin{align}
  A(\boldsymbol{\theta})& = -\frac{h}{2}[2(\text{c}_1+\text{c}_3-2\text{c}_2)\nonumber\\*
  &\quad+(1+2h^2)(\text{s}_1\text{s}_2\text{c}_3+ \text{c}_1\text{s}_2\text{s}_3-2\text{s}_1\text{c}_2\text{s}_3)],\nonumber\\
  B(\boldsymbol{\theta})& = 6h(\text{c}_1+\text{c}_3-2\text{c}_2),\nonumber\\
  C(\boldsymbol{\theta})& = -\frac{h}{2}[\text{s}_1+\text{s}_3-2\text{s}_2\nonumber\\*
  &\quad+(1+2h^2)(\text{s}_1\text{c}_2\text{c}_3+ \text{c}_1\text{c}_2\text{s}_3-2\text{c}_1\text{s}_2\text{c}_3)],\nonumber\\
  D(\boldsymbol{\theta})& = 2h(\text{s}_1+\text{s}_3-2\text{s}_2),
\end{align}
}where we again used notation $\text{s}_i \equiv \sin\theta_i$ and $\text{c}_i \equiv \cos\theta_i$.
The corresponding integrals are given by
{\allowdisplaybreaks
\begin{align}
    I_3[A(\boldsymbol{\theta})]&=\frac{h (-4 h^2+\tau^2 \gamma^2+22) \sin \lambda-12 h \tau \gamma}{2 \gamma^3}\nonumber\\*
    &\quad+\frac{3 (2 h^3+h) \tau \gamma \cos \lambda-2 h (2 h^2+1) \sin (2 \lambda)}{2 \gamma^3},\nonumber\\
    I_3[B(\boldsymbol{\theta})]&=\frac{3 h [12 \tau \gamma-(\tau^2 \gamma^2+24) \sin \lambda]}{\gamma^3},\nonumber\\
    I_3[C(\boldsymbol{\theta})]&=\frac{\tau \gamma (6 (2 h^3+h) \sin \lambda-h \tau \gamma (\cos \lambda+2))}{4 \gamma^3}\nonumber\\*
    &\quad-\frac{16 h \sin ^2\frac{\lambda}{2} [(2 h^2+1) \cos \lambda+4 h^2-1]}{4 \gamma^3},\nonumber\\
    I_3[D(\boldsymbol{\theta})]&=\frac{h[\tau^2 \gamma^2 (\cos \lambda+2)-48 \sin ^2\frac{\lambda}{2}]}{\gamma^3}.
\end{align}}

We find that the new three-body terms improve the agreement with exact numerics in this case but they are not sufficient and one needs to go to an even higher order. We leave this as a future problem.

\section{Global driving}
\label{app:globaldriving}

In this section, we demonstrate a distinguishing feature of the perturbative structure of $H_F$ for local and global driving. The time-independent part ($H_0$) remains the same as before, we only make the driving part spatially uniform given by choosing $V=\gamma \sgn(\sin(\omega t))\sum_{i=1}^NS^x_i$.
The rotating-frame Hamiltonian is now given by
\begin{align}
 H_{rot}(t)&=W^{\dagger} H_0 W\nonumber\\
    &=\sum_{i=1}^{N-1}(-1)^{i}{\bf S}_i\cdot{\bf S}_{i+1}\nonumber\\
    &\quad-h\left(\cos\theta \sum_{i=1}^NS^z_1+\sin\theta \sum_{i=1}^NS^y_1\right).
\end{align}
$W(t)$ gives a global rotation of all the terms about the x-axis. Consequently, the Heisenberg part remains intact because of the $SU(2)$ symmetry. The zeroth term ($l=0$) in the F-M is given by
\begin{align}
 H_F^{(0)}&=\sum_{i=1}^{N-1}(-1)^{i}{\bf S}_i\cdot{\bf S}_{i+1}\nonumber\\
 &\quad-\frac{h}{\lambda}\left[\sin\lambda\sum_{i=1}^N S^z_i+(1-\cos\lambda)\sum_{i=1}^NS^y_i\right],
\end{align}
where $\lambda=\gamma \tau/2$. We get $H_F^{(1)}=0$ as before. 

\begin{figure}
    \centering
    \includegraphics[width=8.6cm]{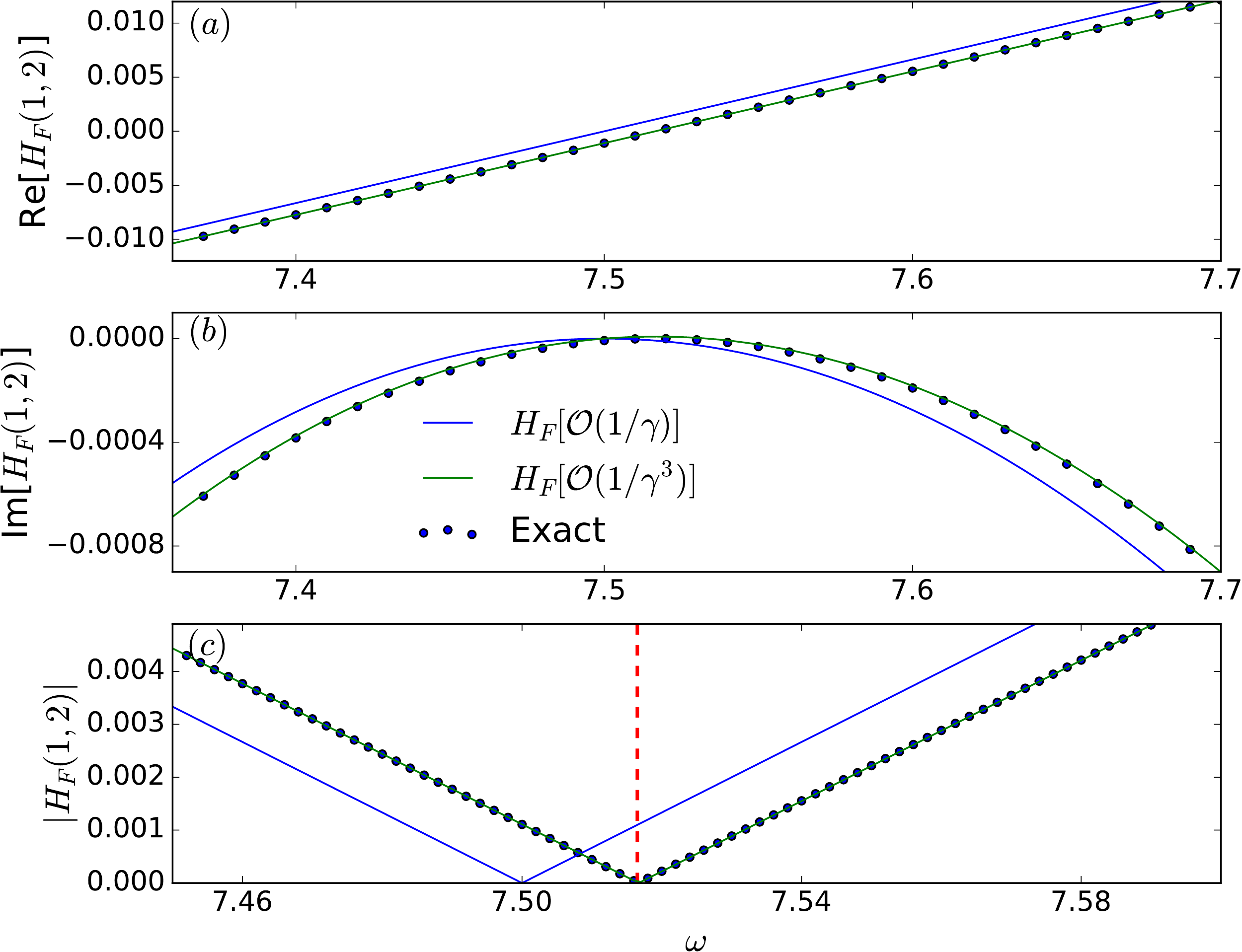}
    \caption{The matrix element $H_F(2,1)$ vs $\omega$ for global driving comparing $H_F^{(2)}$ with exact numerical data. $\gamma=15$, $h=1$, $N=8$. }
    \label{fig:global}
\end{figure}

To calculate the 2nd term ($l=2$) in F-M, we find
\begin{align}
  &\{[H_{rot}(t_1), [H_{rot}(t_2), H_{rot}(t_3)]] + (1\leftrightarrow3)\} = \nonumber\\
  &\quad=-h^3 \left[A(\boldsymbol{\theta}) \sum_{i=1}^N S^z_i + B(\boldsymbol{\theta}) \sum_{i=1}^NS^y_i \right],
\end{align}
where
{\allowdisplaybreaks
\begin{align}
  A(\boldsymbol{\theta}) &= \sin\theta_2\sin(\theta_1+\theta_3) - 2\sin\theta_1\cos\theta_2\sin\theta_3,\nonumber\\
  B(\boldsymbol{\theta}) &= \cos\theta_2\sin(\theta_1+\theta_3) - 2\cos\theta_1\sin\theta_2\cos\theta_3.
\end{align}
}Thus, using the integrals in Eq.~\eqref{integrals2}, we get 
\begin{equation}
  H_F^{(2)} = a(\boldsymbol{p}) \sum_{i=1}^N S^z_i + b(\boldsymbol{p}) \sum_{i=1}^N S^y_i,
\end{equation}
where
{\allowdisplaybreaks
\begin{align}
  a(\boldsymbol{p}) &= \frac{h^3}{6\tau\gamma^3}[2(\sin\lambda+\sin2\lambda) - 3\gamma \tau\cos\lambda],\nonumber\\
  b(\boldsymbol{p}) &= \frac{h^3}{6\tau\gamma^3}[6-2(2\cos\lambda+\cos2\lambda) - 3\gamma \tau\sin\lambda].
\end{align}
}Note that, unlike the local driving, the $l=2$ terms here are at least $\mathcal{O}(1/\gamma^2)$ or smaller. 
Finally, the full $H_F$ can be obtained by adding all these contributions : $H_F[\mathcal{O}(1/\gamma^3)]=H_F^{(0)}+H_F^{(1)}+H_F^{(2)}$. Thus, we see, only the field terms are renormalized but no additional interacting term is generated. Therefore, in this case, the global driving only induces a fully coherent oscillation from the initial x-polarized ($\psi_0$) state at any drive parameter regime. The driven state being a globally rotated version of $\psi_0$ always remains an eigenstate of the $SU(2)$ symmetric Heisenberg interaction part and hence does not suffer any dephasing. In this case, the field terms can get completely canceled at fine-tuned drive frequencies resulting in true freezing of the wavefunction. The exact numerical data including the shift of the freezing frequencies (from $\omega^k_c$ to $\omega^k_m$) can be accurately captured by $H_F[\mathcal{O}(1/\gamma^3)]$ as shown in Fig.~\ref{fig:global}.

\section{Level statistics of the Floquet Hamiltonian}
\label{app:levelstat}
\begin{figure}
    \centering
    \includegraphics[width=\linewidth]{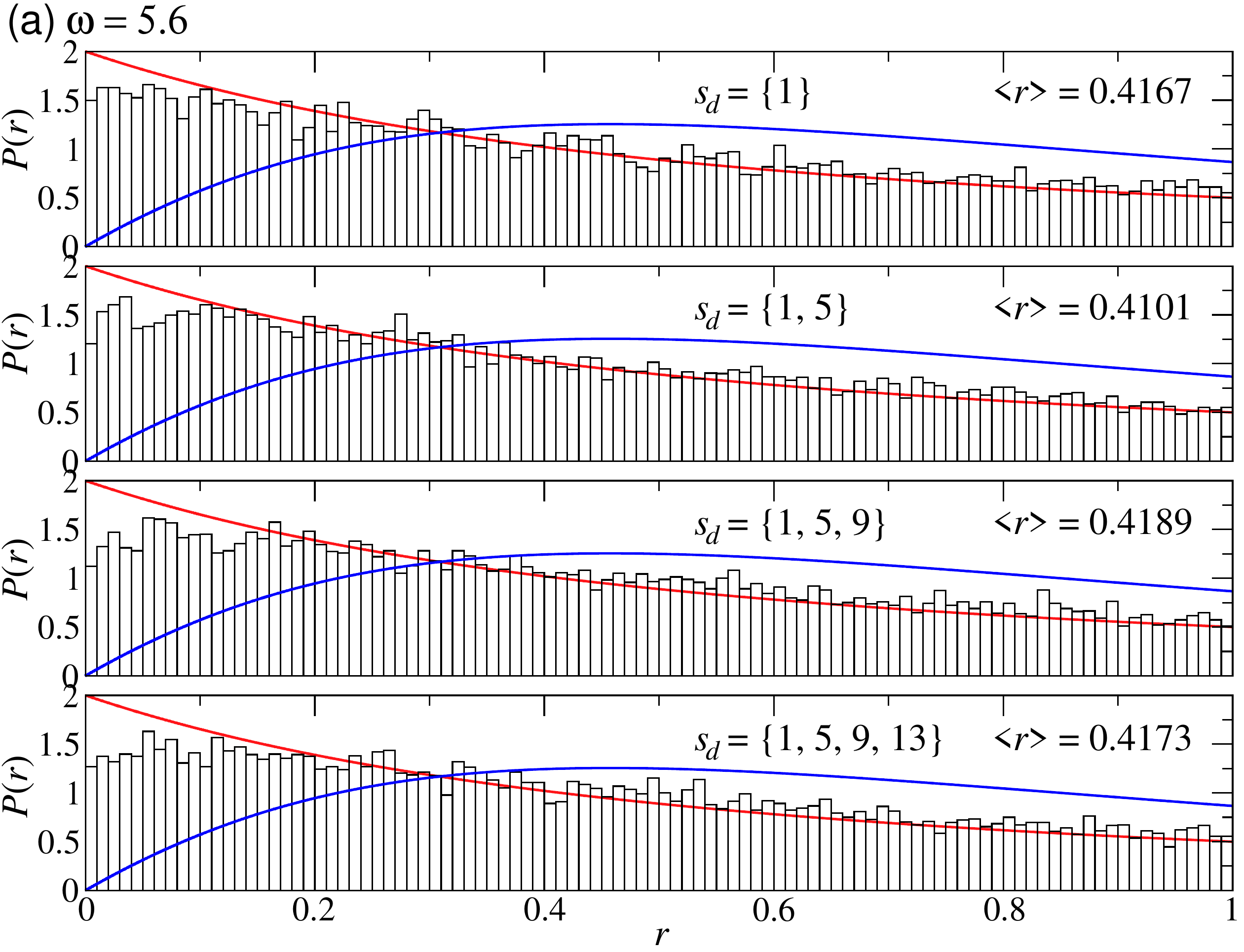}\\
    \includegraphics[width=\linewidth]{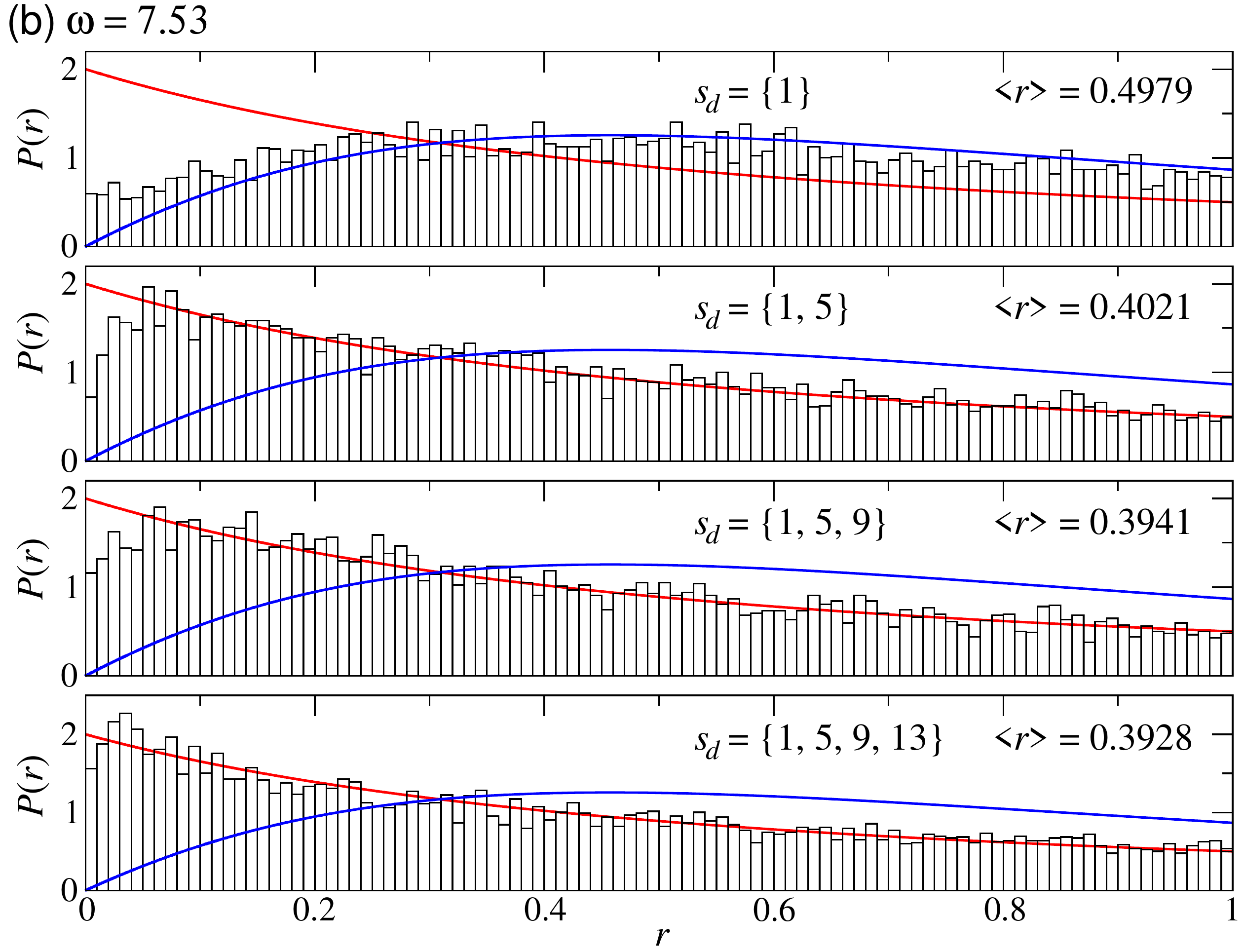}
    \caption{Level statistics of the Floquet eigenvalues, (a)~at a generic drive frequency $\omega=5.6$ (b)~at the freezing frequency $\omega=\omega^1_m=7.53$. The blue line corresponds to the Wigner surmise (an approximation to the GOE statistics of chaotic systems), while the red line is the Poisson distribution (integrable systems). $N=14$, $\gamma=15$.}
    \label{fig:r}
\end{figure}

In this appendix, we study the level statistics of the Floquet Hamiltonian ($H_F$). At generic drive frequencies ($\omega\lesssim \gamma$) $H_F$ hardly has any symmetry and hence we increasingly order the Floquet eigenvalues ($\epsilon_i$) in the entire first Floquet BZ. At the freezing frequencies, $D_z$ is an emergent conserved quantity and we confine ourselves in a specific $D_z$ sector. We then calculate the gap between consecutive eigenvalues: $g_i=\epsilon_{i+1}-\epsilon_i$ (we are using a different symbol for the gap to distinguish it from the spectral gap defined in the main text) and their ratios defined as 
\begin{equation}
    r_i=\frac{\text{min}(g_i,g_{i+1})}{\text{max}(g_i,g_{i+1})}.
\end{equation}
The distribution of $r_i$ gives important information about the integrability of the system.
For example, chaotic (nonintegrable) systems exhibit a GOE distribution, whereas integrable systems admit Poisson level statistics. In our case, we find the distribution to be more like Poissonian at generic drive frequencies (see Fig.~\ref{fig:r}). Here we note that in addition to integrability (an unlikely possibility in our case), Poissonian statistics can appear in several other situations, for example, if there is some hidden unresolved symmetry or if the Hilbert space is fragmented. We find the following operator to be a good candidate of an approximate discrete symmetry at any frequency 
\begin{equation}
    \mathcal{M} = 2^N \prod_{i\in s_d}\left[\left|\cos{\lambda\over2}\right| \left(S^z_1+\tan{\lambda\over2}S^y_1\right)\right]\prod_{i\notin s_d} S^z_i,
\end{equation}
with eigenvalues $\pm1$. In fact, $\mathcal{M}$ is conserved exactly at the level of $H_F^{(0)}$ for $s_d=\{1\}$, i.e.\@ $[\mathcal{M}, H_F^{(0)}] = 0$ for any $\lambda$. Note that $\mathcal{M}$ converts to $D_z$ when $\lambda = 2\pi k$. We also numerically find that for any driving protocol with every fourth site driven, $\langle \Phi_r | \mathcal{M} | \Phi_r \rangle \approx \pm 1$ on the full range of $\lambda$. The existence of such operators, which may remain weakly conserved, is most likely the origin of the Poissonian nature of the level statistics at generic drive frequencies.
At the freezing frequency, we find the distribution to resemble GOE for $s_d=\{1\}$ but again Poissonian for protocols $s_d=\{1,5\}, \{1,5,9\}, \{1,5,9,13\}$.  

\section{More results on the emergence of SZM and entanglement of the Floquet eigenstates}
\label{app:SZMmore}

Here we corroborate the emergence of SZM (presented in the main text) with more numerical results. If a Floquet system described by the Floquet Hamiltonian $H_F$ possesses a SZM with $S^x_1$ as the leading part, then the whole Floquet spectrum can be decomposed into quasi-degenerate eigenpairs $|\Phi^{\pm}_r\rangle$ (with $D^z|\Phi^{\pm}_r\rangle = \pm|\Phi^{\pm}_r\rangle$ and $H_F|\Phi^{\pm}_r\rangle = \epsilon^{\pm}_r |\Phi^{\pm}_r\rangle$) where $|\Phi^{\pm}_r\rangle$ can be obtained from $|\Phi^{\mp}_r\rangle$ under the action of $S^x_1$ up to a global phase.
We increasingly order both $\epsilon^+_r$ and $\epsilon^-_r$s  within the first Floquet-BZ, and also define $\langle \xi^-_r|\xi^+_r\rangle = \langle\bar{\xi}^-_r|\bar{\xi}^+_r\rangle = \mathcal{R}_re^{i\theta_r}$, with $\mathcal{R}_r < 1$ being the absolute value and $\theta_r$ being the phase. Such labeling of eigenpairs may lead to some subtleties in some cases which we clarify now. 
We find that some Floquet eigenpairs (say, $\epsilon^\pm_p$), obtained in this way, may yield a super low value of $|\langle\Phi_p^-|S^x_1|\Phi_p^+\rangle|$ instead of $\sim0.5$. Such ghost eigenpairs are found to appear always in pairs (say, $p,p+1$) and a cross pairing may restore the value of the matrix element i.e.\@ $|\langle\Phi_{p(p+1)}^-|S^x_1|\Phi_{p+1,(p)}^+\rangle|\sim0.5$. Such cross-pairing may also reduce one of the pairing gaps but then the other one will definitely increase. 
This kind of ghost pairs get rarer with the emergence of SZM. But, when one of the Floquet eigenvalues of a normal pair crosses the Floquet BZ boundary, that results in a cascade of mismatch between the increasingly ordered $\epsilon^\pm_r$ which generates ghost pairs throughout the spectrum. This leads to a sharp drop in $\overline{|\langle\Phi^-|S^x_1|\Phi^+\rangle|}$ at some $\omega$ which are mostly away from the special frequencies ($\omega^k_m$) where even the conservation of $D_z$, the first criteria of having a SZM, doesn't satisfy. 
We discard such rare pathological points from our consideration while plotting Fig.~\ref{fig:szm}(b) in the main text. 
Thus the correspondence between the increasingly ordered $\epsilon^{\pm}_r$ may get ill-defined at drive frequencies away from the slowest relaxation points ($\omega^k_m$). But as we increase the number of optimally chosen driven sites ($n_d^\text{4th}$) and tune the drive frequency towards $\omega_m$, the correspondence gradually becomes more and more accurate signaling the emergence of SZM.

\begin{figure*}
\begin{center}
 \includegraphics[width=0.9\linewidth]{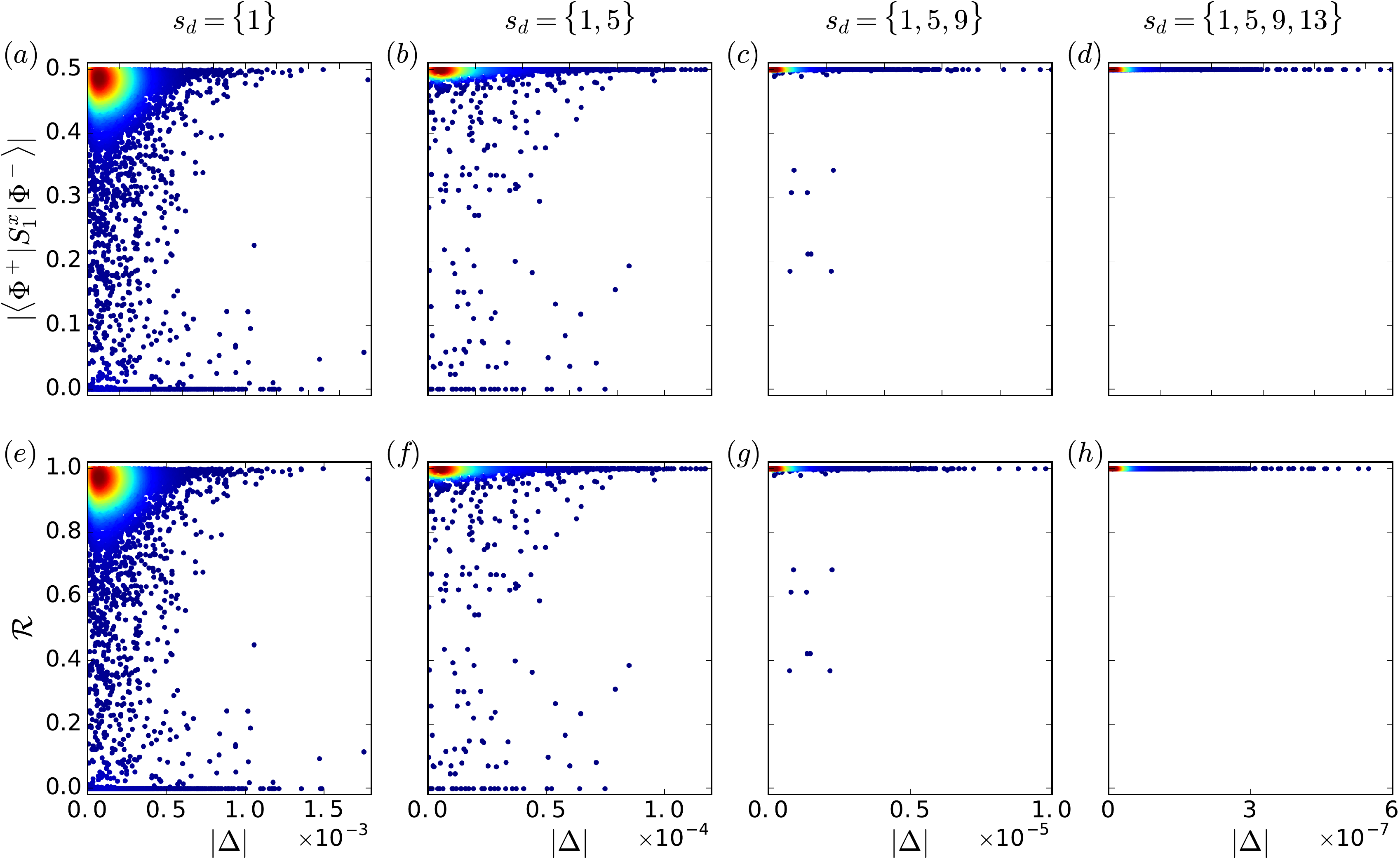}
 \caption{The behavior of $|\langle\Phi_r^-|S^x_1|\Phi_r^+\rangle|$ in the upper panels (a-d) in parallel with the behaviors of $\mathcal{R}_r$ in the lower panels (e-h). The driven sites are mentioned on top of the upper panels (valid for both panels in each column).
 The values of $\overline{|\langle\Phi^-|S^x_1|\Phi^+\rangle|}$ are (a)~0.4275, (b)~0.4942, (c)~0.4996, (d)~0.4998 and the values of $\overline{\mathcal{R}}$ are (e)~0.8550, (f)~0.9883, (g)~0.9991, (h)~0.9996. $\gamma=15$, $\omega=7.53$, $N=14$ for all the plots.}
 \label{fig:sxmn}
 \end{center}
\end{figure*}

We demonstrate this in Fig.~\ref{fig:sxmn} where we show the behavior of $|\langle\Phi_r^-|S^x_1|\Phi_r^+\rangle|$ in parallel with the behaviors of $\mathcal{R}_r$ with increasing $n_d^\text{4th}$.
One can see that there are many pairs with $|\langle\Phi_r^-|S^x_1|\Phi_r^+\rangle|<0.5$ including the ghost pairs with $|\langle\Phi_r^-|S^x_1|\Phi_r^+\rangle|\sim 0$ for $s_d=\{1\}$.
The number of such pairs decreases with increasing $n_d^\text{4th}$ and totally disappears when every fourth site is driven in a chain of fixed length $N$ [as shown in Fig.~\ref{fig:sxmn}(d,h)].
This follows from the consistent behavior of $\mathcal{R}$ (as shown in Fig.~\ref{fig:sxmn} (e)-(h)) which was conjectured in the main text as a requirement for the emergence of SZM. 
We note here that the lower value of $\overline{|\langle\Phi^-|S^x_1|\Phi^+\rangle|}$ for $s_d=\{1\}$ compared to the value for other higher $n_d^\text{4th}$ is not an artifact of just the presence of ghost pairs.
There are many pairs with $|\langle\Phi^-|S^x_1|\Phi^+\rangle|$ a bit less than 0.5 for direct pairing albeit much higher than the corresponding value for cross pairing. Such pairs also reduce in number and disappear with increasing $n_d^\text{4th}$, ramping up the value of $\overline{|\langle\Phi^-|S^x_1|\Phi^+\rangle|}$ towards 0.5. The parallel approach of all the $\mathcal{R}_r$ towards one enable us to arrive at the condition: $S^x_1\ket{\Phi^{\pm}_r}=\frac{1}{2}e^{\pm i\theta_r}\ket{\Phi^{\mp}_r}$ via the ansatz used in Eq.~\ref{eq:Floquet_eigenstates} in the main text.

\begin{figure*}
\centering
\includegraphics[width=0.9\linewidth]{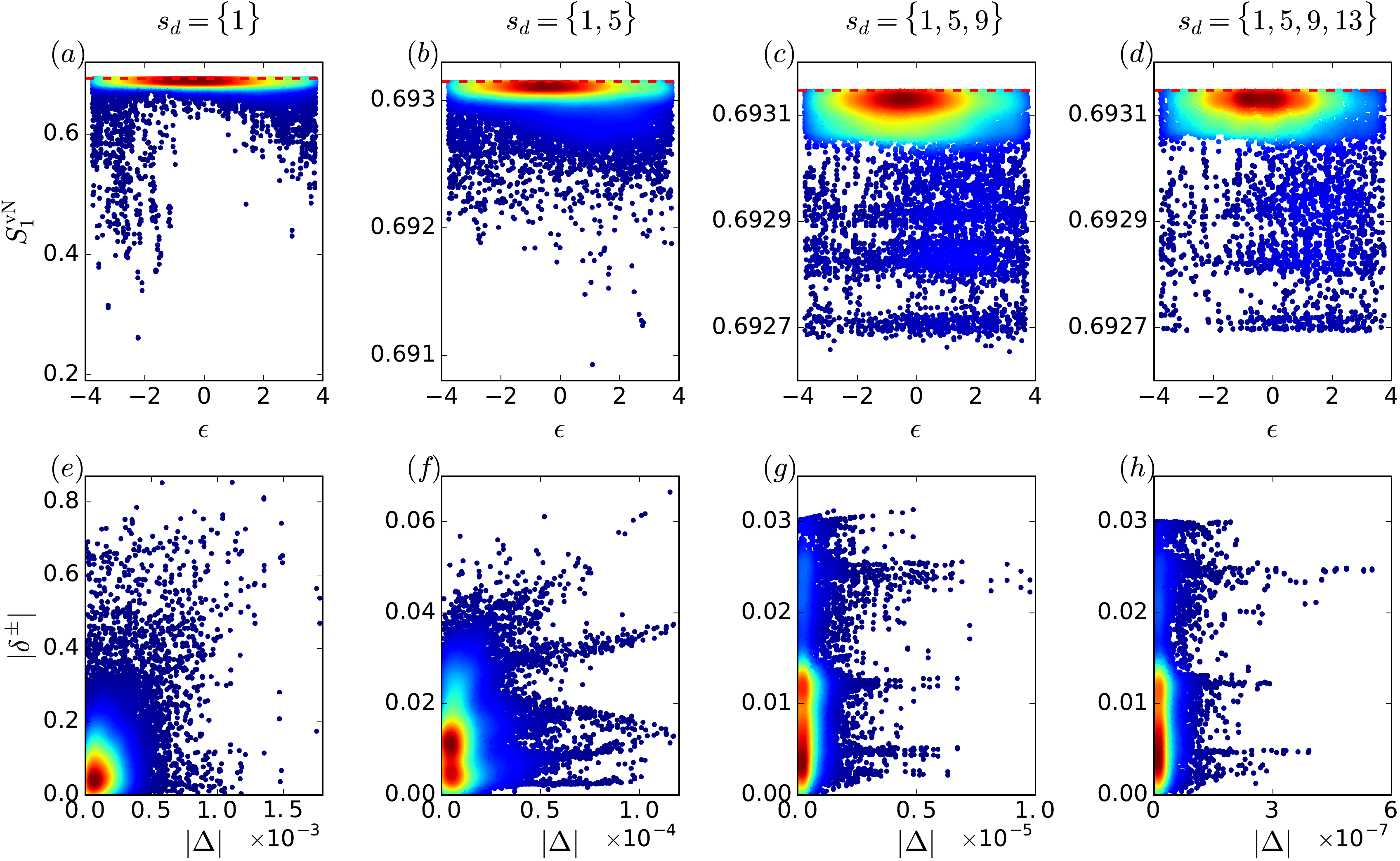}
\caption{The behavior of $S^{\text{vN}}_1$ in all the Floquet eigenstates in the upper panels in parallel with the behavior of $|\delta^{\pm}|$ (not distinguished) in the lower panels. The average deviations of $S_1$ from the maximal value $\ln 2$ (shown in red dashed line) are (a)~$1.21\times10^{-2}$, (b)~$1.42\times10^{-4}$, (c)~$8.89\times10^{-5}$, (d)~$8.34\times10^{-5}$, and the average values of $|\delta^{\pm}|$ are (e)~0.11, (f)~$1.38\times10^{-2}$, (g)~$1.09\times10^{-2}$, and (h)~$1.05\times10^{-2}$. All other parameters are the same as in Fig.~\ref{fig:sxmn}.}
\label{fig:Ent_delta}
\end{figure*}

In Fig.~\ref{fig:Ent_delta} we show the behavior of $S^{\text{vN}}_1$ for all the Floquet eigenstates in parallel with the behaviors of $\delta^{\pm}_r=\langle \xi^{\pm}_r|\bar{\xi}^{\pm}_r\rangle$. 
There are lots of low-entangled Floquet eigenstates for $s_d=\{1\}$ but all of them get strongly entangled for $s_d=\{1,5\}$. This is accompanied by a huge reduction in the average value of $|\delta^{\pm}|$, we denote by $\overline{|\delta^{\pm}|}$. Further increase of driven sites ($n_d^\text{4th}$) is found to reduce $\overline{|\delta^{\pm}|}$ only a little bit. 
Consequently, the entanglement spectrum also becomes only slightly narrower, though by now they are already saturated at their maximal value to a good extent. Thus in the asymptotic limit, when every fourth site is driven, the boundary spin forms Bell like pairing with the rest of the system in all the Floquet eigenstates.

\section{Semi-analytical study of the pairing-gap}
\label{app:semianalytical_gap}

We now seek to have some more understanding of the vanishing of the pairing gap which is given by
\begin{align}
\Delta_r&=\langle \Phi^+_r|H_F|\Phi^+_r\rangle-\langle \Phi^-_r|H_F|\Phi^-_r\rangle\nonumber\\
&=\frac{1}{2}\left[(\langle{\rightarrow},\xi^+_r|H_F|{\rightarrow},\xi^+_r\rangle-\langle{\rightarrow},\xi^-_r|H_F|{\rightarrow},\xi^-_r\rangle)\vphantom{\frac{}{}}\right.\nonumber\\
&\quad\left.+\vphantom{\frac{}{}}(\langle{\leftarrow},\bar{\xi}^+_r|H_F|{\leftarrow},\bar{\xi}^+_r\rangle-\langle{\leftarrow},\bar{\xi}^-_r|H_F|{\leftarrow},\bar{\xi}^-_r\rangle)\right]\nonumber\\
&\quad+\text{Re}[\langle {\rightarrow},\xi^+_r|H_F|{\leftarrow},\bar{\xi}^+_r\rangle]+\text{Re}[\langle {\rightarrow},\xi^-_r|H_F|{\leftarrow},\bar{\xi}^-_r\rangle],
\label{eq:gap}
\end{align}
%where $|s^x_1,\psi\rangle=|s^x_1\rangle\otimes|\psi\rangle$, $|s^x_1\rangle$ and $|\psi\rangle$ represent eigenstates of $S^x_1$ and states of the rest of the system (excluding the left boundary site) respectively.

The quantities inside the first parenthesis in the second line of Eq.~\ref{eq:gap} get canceled with the emergence of SZM due to the condition $\mathcal{R}_r\approx 1$, which we have verified numerically.
The vanishing of the quantities in the third line of Eq.~\ref{eq:gap} (let us call it $\Delta'_r$) is less straightforward to see and here we will adapt a semi-analytical approach to understand this. To this end, we first decompose $H_F$ into three parts: $H_F = h^F_A + h^F_{A,B} + h^F_{B}$, where the subscripts denote part of the chain (sites) on which each term is supported: $A$ and $B$ represent site 1 and rest of the chain respectively, whereas $h^F_{A,B}$ represent terms that connect these two parts,
\begin{align}
  h^F_{A,B}&=-S^x_1S^x_2 \nonumber\\
  & - \left[\left(\frac{\sin\lambda}{\lambda}-a_4(\boldsymbol{p})\right) S^y_1S^y_2 + \left(\frac{\sin\lambda}{\lambda} - a_3(\boldsymbol{p})\right)S^z_1S^z_2\right]\nonumber\\
  & + \left[\left(\frac{1-\cos\lambda}{\lambda}+a_5(\boldsymbol{p})\right)S^z_1S^y_2\right. \nonumber\\
  & \left. - \left(\frac{1-\cos\lambda}{\lambda} - a_6(\boldsymbol{p})\right)S^y_1S^z_2\right] + H^{(2)}_{F,\text{nonloc}}.
\end{align}
The purpose of such decomposition of the Floquet Hamiltonian together with the eigenstates as in Eq.~\ref{eq:Floquet_eigenstates} will be clear now. We first note that $\langle{\rightarrow} | h^F_A |{\leftarrow}\rangle$ is nothing but $H_F(2,1)$ as discussed in Sec.~\ref{sec:singlesite}. Secondly, note that though we have quite accurate knowledge of $h^F_A$ and $h^F_{A,B}$, the analytical structure of $h^F_B$ is not very well known particularly for multisite driving. The former also remains almost unchanged as we increase $n_d^\text{4th}$ but the latter must change drastically with it. But $\langle{\rightarrow}, \xi^{\pm}_r| h^F_B | {\leftarrow}, \bar{\xi}^{\pm}_r \rangle = 0$ and hence $h^F_B$ has zero contribution in $\Delta'_r$.
This enables us to write down a semi-analytical expression of $\Delta'_r$,
\begin{align}
 \Delta'_r
  &= \text{Re}[\bra{\rightarrow} h^F_A \ket{\leftarrow} (\delta^+_r+\delta^-_r)+\langle \rightarrow,\xi^+_r|h^F_{A,B}|\leftarrow,\bar{\xi}^+_r\rangle\nonumber\\
  &\quad+\langle \rightarrow,\xi^-_r|h^F_{A,B}|\leftarrow,\bar{\xi}^-_r\rangle].
  \label{eq:semianalytic}
\end{align}

\begin{figure}
    \centering
    \includegraphics[width=0.99\linewidth]{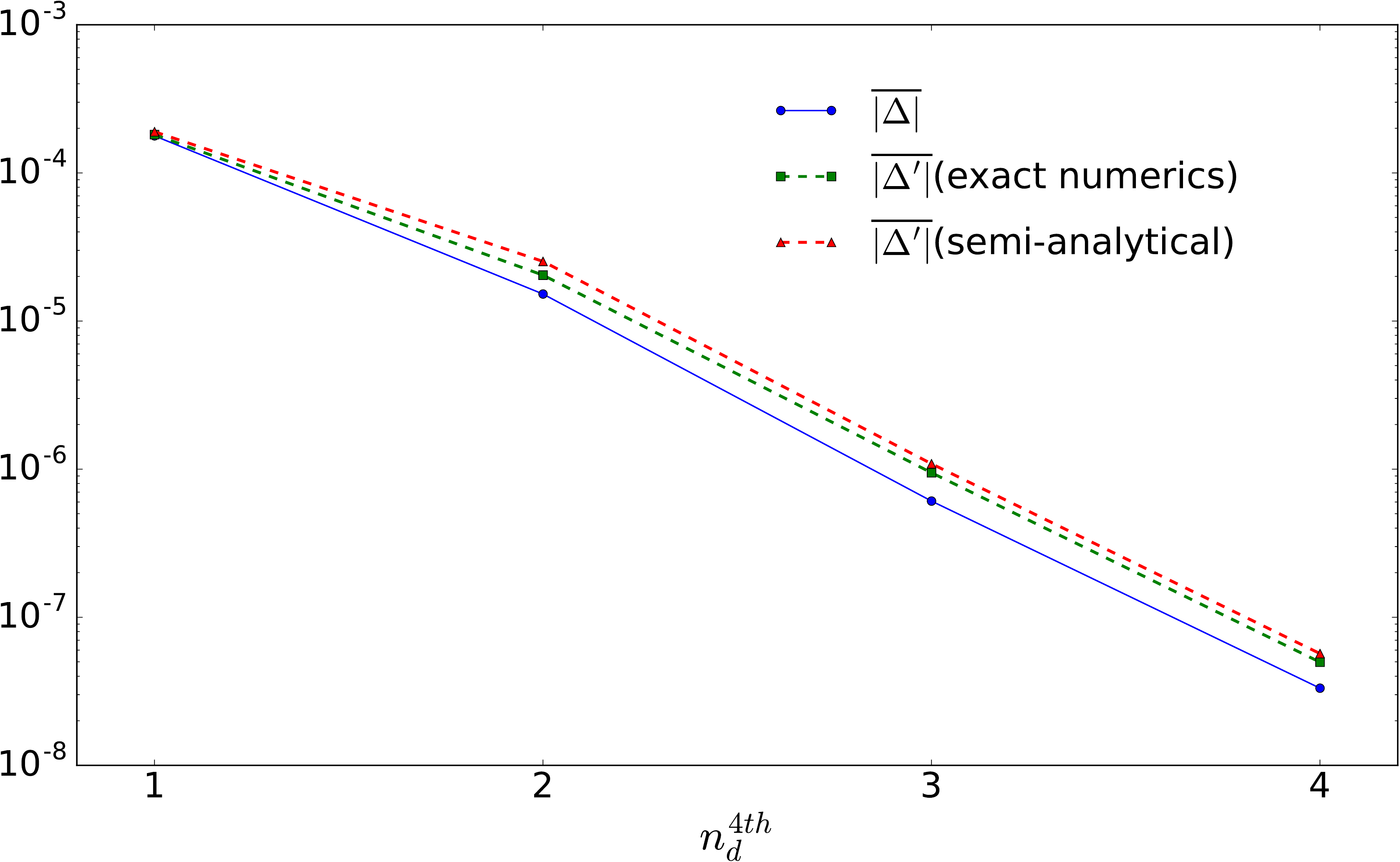}
    \caption{Comparison of the scaling of $\overline{|\Delta|}$ and $\overline{|\Delta'|}$. The latter is calculated from exact numerics as well as using a semi-analytical expression (Eq.~\ref{eq:semianalytic}). Parameters: $h=1,\gamma=15,\omega=\omega^1_m=7.53,N=14$.} 
    \label{fig:semianalytical_gap}
\end{figure}

$\bra{\rightarrow} h^F_A \ket{\leftarrow}$ becomes very small at $\omega \approx \omega_m$ whereas $|\delta^{\pm}|$ get smaller with increasing $n_d^\text{4th}$. We found in Appendix~\ref{app:SZMmore} that $\overline{|\delta^{\pm}|}$ reduces strongly when we go from $n_d^\text{4th}=1$ ($s_d=\{1\}$) to $n_d^\text{4th}=2$ ($s_d=\{1,5\}$) but further increase in $n_d^\text{4th}$ does not reduce the value of $\overline{|\delta^{\pm}|}$ much (Fig.~\ref{fig:Ent_delta}). Interestingly, the contribution to the gap ($\Delta$) which comes only from the real part of these complex numbers, gets significantly smaller due to some cancellations between $\delta^{\pm}$. We compare $\overline{|\Delta'|}$ from Eq.~\ref{eq:semianalytic} with fully exact numerics (using the final line of Eq.~\ref{eq:gap}) in Fig.~\ref{fig:semianalytical_gap} and find very good agreement. Moreover, we find, $\overline{|\Delta'|}$ itself is not much different than $\overline{|\Delta|}$. This again implies that our understanding of the structure of the Floquet Hamiltonian is quite accurate.

\bibliography{refs}
\end{document}